\documentclass[twoside,romanappendices,twocolumn]{IEEEtran} %

\usepackage[pdftex]{graphicx}
\graphicspath{}
\DeclareGraphicsExtensions{.pdf}
\usepackage{amsmath}
\usepackage[caption=false,font=footnotesize]{subfig}

\usepackage{amssymb}

\newcommand{\R}{\mathbb{R}}
\newcommand{\N}{\mathbb{N}}

\newcommand{\E}[1]{\mathrm{E} \left[ #1 \right] }
\newcommand{\Ei}[1]{\mathrm{E} [ #1 ] }
\newcommand{\condE}[2]{\mathrm{E} \left[ #1 \middle| #2 \right] }
\newcommand{\I}[3]{I_{#1}\left(#2,#3\right)}
\newcommand{\Ii}[3]{I_{#1}(#2,#3)}
\newcommand{\Var}{\mathrm{Var}}

\newcommand{\SNR}{\mathrm{SNR}}
\newcommand{\dB}{\text{dB}}
\newcommand{\prob}[1]{\mathbb{P} \left( #1 \right)}
\newcommand{\probi}[1]{\mathbb{P} ( #1 )}
\newcommand{\condProb}[2]{\mathbb{P} \left( #1 \middle| #2 \right)}
\newcommand{\proj}[2]{\text{P}_{#2}{#1}}
\newcommand{\entropy}[1]{\mathcal{S}\left(#1\right)}
\newcommand{\entropyi}[1]{\mathcal{S}(#1)}
\newcommand{\relent}[2]{\mathcal{H}(#1,#2)}
\newcommand{\sparsity}{s}
\newcommand{\overcomp}{o}
\newcommand{\overcompD}{o^{\dagger}}
\newcommand{\overcompS}{o^{*}}
\newcommand{\ball}[2]{\mathcal{B}_{#2}(#1)}
\newcommand{\area}[1]{\mathcal{A}(#1)}
\newcommand{\myBeta}[2]{\text{Beta}\left(#1,#2\right)}
\newcommand{\myBetai}[2]{\text{Beta}(#1,#2)}
\newcommand{\optSuccProb}[1]{ \mathcal{P} (#1) }
\newcommand{\optSuccProbWC}{ \mathcal{P}^* }

\newcommand{\bigtheta}[1]{ \Theta \left(#1\right) }
\newcommand{\bigchi}{\mbox{\Large$\chi$}}% big chi
\newcommand{\norm}[1]{\left\lVert #1 \right\rVert}
\newcommand{\normi}[1]{\| #1 \|}
\newcommand{\ceil}[1]{\left\lceil #1 \right\rceil}

\newcommand{\eg}{\emph{e.g.}, }
\newcommand{\ie}{\emph{i.e.}, }

\renewcommand{\IEEEQED}{\IEEEQEDopen}

\begin{document}

\title{On the Minimal Overcompleteness Allowing Universal Sparse Representation}


\author{Rotem~Mulayoff and Tomer~Michaeli%
\thanks{This research was partly supported by the Ollendorff Foundation, by an Alon Fellowship, and by the Israel Science Foundation (grant no.~852/17).}%
\thanks{R.~Mulayoff and T.~Michaeli are with the Department of Electrical Engineering, Technion–-Israel Institute of Technology, Haifa 32000, Israel (e-mails: smulayof@campus.technion.ac.il, tomer.m@ee.technion.ac.il).}}% <-this % stops a space


\markboth{IEEE TRANSACTIONS ON INFORMATION THEORY}{Mulayoff and Michaeli: On the Minimal Overcompleteness Allowing Universal Sparse Representation}

\IEEEpubid{\begin{minipage}{\textwidth}\ \\[12pt]
	\begin{center}
		Copyright~\copyright~2017 IEEE. Personal use of this material is permitted.\\
		However, permission to use this material for any other purposes must be obtained from the IEEE by sending a request to pubs-permissions@ieee.org.
	\end{center}
\end{minipage}}

% Make the title area
\maketitle

\begin{abstract}
Sparse representation over redundant dictionaries constitutes a good model for many classes of signals (\eg patches of natural images, segments of speech signals, etc.). However, despite its popularity, very little is known about the representation capacity of this model. In this paper, we study how redundant a dictionary must be so as to allow \emph{any vector} to admit a sparse approximation with a prescribed sparsity and a prescribed level of accuracy. We address this problem both in a worst-case setting and in an average-case one. For each scenario we derive lower and upper bounds on the minimal required overcompleteness. Our bounds have simple closed-form expressions that allow to easily deduce the asymptotic behavior in large dimensions. In particular, we find that the required overcompleteness grows exponentially with the sparsity level and polynomially with the allowed representation error. This implies that universal sparse representation is practical only at moderate sparsity levels, but can be achieved with a relatively high accuracy. As a side effect of our analysis, we obtain a tight lower bound on the regularized incomplete beta function, which may be interesting in its own right. We illustrate the validity of our results through numerical simulations, which support our findings.

\end{abstract}


% Counters for theorems numbering
\newcounter{Definitions}
\newcounter{Lemmas}
\newcounter{Theorems}
\newcounter{Corollaries}

\section{Introduction} \label{myIntro}
\IEEEPARstart{R}{esearchers} and engineers often use transforms to analyze and process signals. A common desired property from a transform, is that it allow signals to be represented as combinations of a small number of ``atoms''. For example, the Fourier transform is commonly used for analyzing audio signals \cite{gold2011speech} since they tend to be comprised of a small number of harmonic components. Piecewise smooth signals, on the other hand, are much more compactly represented by the Wavelet transform, which is thus popular in image processing \cite{chan2005image}. The emergence of the field of sparse representations \cite{mallat1993matching}, initiated the systematic construction of dictionaries, that allow representing signals as linear combinations of a \emph{small} number of their atoms \cite{engan1999method,aharon2006rm}. Today, this concept constitutes a key ingredient in numerous areas, ranging from image enhancement to signal recovery and compression~\cite{elad2010sparse,eldar2012compressed,zhang2015survey}.

In this paper, we address a fundamental question relating to the expressive power of sparse representations. Specifically, we study conditions under which a redundant dictionary can be used to represent every signal in \(\R^d\) as a linear combination of at most \(k<d\) of its atoms, with an error no larger than \( \varepsilon \). Our goal is to obtain necessary and sufficient conditions on the minimal number of atoms~$n$ allowing this.

This problem has two motivations. First, when the sparse representation model is used as a prior, as in compressed sensing or signal restoration \cite{elad2006image,yang2010image}, only a small set of signals is meant to be sparsely representable over the dictionary. This is often achieved by learning a dictionary from a set of relevant training examples (\eg patches from natural images) \cite{engan1999method,aharon2006rm,lewicki2006learning,mairal2009online}. In this context, it is of interest to identify when a dictionary has an unnecessarily large overcompleteness (\ie one which allows sparse representation of every signal, and not only of those from the designated set). A second motivation relates to the use of the sparse representation model as a generic transform, under which all signals are sparse.

%In the vast majority of applications that use sparse representations, only a small set of signals is meant to be sparsely representable over the dictionary. This is often achieved by learning a dictionary from a set of relevant training examples (\eg patches from natural images) \cite{engan1999method,aharon2006rm,lewicki2006learning,mairal2009online}. Such an approach is therefore useful when sparsity is used as a prior for inverse problems, like in compressed sensing or in signal enhancement (denoising, deblurring, etc.) \cite{elad2006image,yang2010image}. However, it is not adequate as a generic tool for signal analysis, because it does not serve as a good transform for signals outside the designated set.

%In this paper, we address a fundamental question relating to the use of sparse representations as a universal tool for signal analysis. Specifically, we study the existence of redundant dictionaries such that every signal in \(\R^d\) can be represented as a linear combination of at most \(k<d\) of their atoms, with an error no larger than \( \varepsilon \). Our goal is to obtain necessary and sufficient conditions on the minimal number of atoms~$n$ allowing this.

It is easy to show that when $k$ is taken to be a fixed fraction of $d$, the set of signals in \(\R^d\) that can be approximately represented by a specific choice of \(k\) atoms has a volume which is exponentially small in $k$ (see Sec.~\ref{Derivations} for details). This is, in fact, the principle underlying the Johnson-Lindenstrauss lemma \cite{johnson1984extensions}. This lemma asserts that when projecting points in $\R^d$ onto a random $k$-dimensional space, there is a concentration of measure effect whereby the points' norms are approximately preserved up to a factor of $k/d$. Therefore, when $k$ is much smaller than $d$, such projections are very far from the original points with high probability. Nevertheless, in our context, if $k$ is also a fixed fraction of $n$, then the number of choices of \(k\) atoms from a dictionary of size \(n\) is exponentially large in $k$. Therefore, it is not a-priori clear whether the overcompleteness should be very large in order to allow universal sparse representation. Interestingly, our results show that for certain regimes of error and sparsity levels, universal sparse representation can be achieved with moderate redundancy. For other regimes, on the other hand, universal sparse representation becomes impractical.

\IEEEpubidadjcol
It should be noted that our setting is very different from that of compressed sensing. There, the quantity of interest is the minimal number of linear measurements from which any $k$-sparse signal can be uniquely recovered \cite{donoho2006compressed,tropp2007signal,donoho2006stable,candes2006stable}. Merging the measurement matrix into the dictionary, this problem is equivalent to asking what is the minimal number of rows of a dictionary allowing to uniquely recover any signal that is $k$-sparse in the standard (non-overcomplete) basis. In contrast, here we analyze the minimal number of atoms (columns of an overcomplete dictionary) with which all signals possess a sparse representation. Furthermore, we do not require uniqueness of the representation.

Mathematically, universal sparse representation can be viewed as a covering problem, a branch of mathematics with many known results \cite{boroczky2004finite}. However most works consider ball covering problems, whereas our setting is concerned with covering by dilations of linear subspaces (spanned by subsets of atoms from the dictionary). Moreover, these subspaces share atoms and are thus constrained to intersect. To the best of our knowledge, such settings were not studied in the past.

Only a few attempts were made to characterize the representation ability of overcomplete dictionaries. In \cite[Ch.~7]{aharonovercomplete} the author provided approximations for the relative volume of signals that admit a sparse representation with an allowable error. However, the expressions depend on properties of the dictionary (minimal and maximal singular values of any subset of columns) and are thus not universal. Furthermore, the accuracy of the approximations in high dimensions is not clear. In \cite{akccakaya2008frame} the authors analyzed a stochastic setting, for which they provided a lower bound on the achievable mean squared error (MSE) of the representation as a function of the sparsity and the dictionary's overcompleteness. The analysis is universal in that it holds for all dictionaries. However, it does not provide an upper bound on the error (from which an upper bound on the required overcompleteness could be deduced), and it does not provide deterministic (worst-case) results.

In this paper we study the universal sparse representation problem from both a worst case standpoint and an average case one. In the worst case setting we request that the representation error be bounded by $\varepsilon$ for every signal in \(\R^d\). In the average case setting, we assume that the signal is random and require that the probability that it can be sparsely represented with an error less than $\varepsilon$, be high. For each scenario we give lower and upper bounds on the minimal required overcompleteness allowing universal sparse representation. As opposed to previous works, our bounds have simple closed-form expressions, which allow to easily deduce the asymptotic behavior of the required overcompleteness. In particular, our bounds reveal that if $\varepsilon \ll 1$ or $k\ll d$, then the minimal required overcompleteness behaves like $ ( 1/ \varepsilon) ^{d/k-1} $ up to polynomial factors in $d/k$. We provide simulations, which show that our bounds correctly predict the threshold at which sparse coding techniques start to succeed in approximating arbitrary signals. As a side effect of our derivations, we obtain a tight lower bound on the regularized incomplete beta function, which may be interesting in its own right.

The paper is organized as follows. Section \ref{MainResults} introduces our problem in mathematical terms, and describes our main results. Section \ref{Derivations} includes the derivations of the bounds, and Section \ref{NumericalComp} presents some numerical experiments, which illustrate and validate the theorems.

\section{Main Results} \label{MainResults}
Our goal is to be able to represent every signal $x\in\R^d$ as a linear combination of a small number $k<d$ of atoms from some dictionary $\Phi\in \R^{d\times n}$. Note that this is impossible to do without incurring some error, since the set of signals admitting such a $k$-sparse representation is a union of $\binom{n}{k}$ subspaces of dimension at most $k$, which is strictly contained in $\R^d$. However, the question we ask is: Under what conditions can we guarantee a $k$-sparse representation for every signal in $\R^d$ \emph{with a small error}?

\newtheorem{RepresentationError}[Definitions]{Definition}
\begin{RepresentationError}[Normalized \(k\)-sparse representation error] \label{RepresentationErrorDef}
	We define the \emph{\(k\)-sparse representation error} of a signal \(x \in \R^d \) over a dictionary \(\Phi \in \R^{d\times n} \) as
	\begin{equation}\label{RepresentationErrorEq}
	\epsilon(x,\Phi) \triangleq \min_{\alpha \in \R^n} \frac{\norm{x-\Phi \alpha} }{\norm{x}} \quad \text{s.t.} \quad \norm{\alpha}_0 \leq k,
	\end{equation}
where the $\ell_0$ (pseudo) norm $\|\cdot\|_0$ counts the number of nonzero elements of its vector argument. We shall say that \(x\) has a \emph{\(k\)-sparse representation over \(\Phi\) with precision \(\varepsilon\)} if \(\epsilon(x,\Phi)\leq \varepsilon\).
\end{RepresentationError}

The normalized error is indifferent to scaling of $x$. Therefore, without loss of generality, we will restrict our analysis to signals lying on the unit sphere (\textit{i.e.}, with \(\norm{x} = 1\)). We are interested in the existence of dictionaries $\Phi$ such that $\epsilon(x,\Phi)$ is small for all, or at least most, signals $x$. More specifically, we consider both a worst-case design (Sec. \ref{WorstCaseResults}) and an average-case one (Sec. \ref{AverageCaseResults}). In the former, we require that the error $\epsilon(x,\Phi)$ be bounded by $\varepsilon$ for every $x\in\R^d$. In the latter, we assume that $x$ is a random vector and require that the probability that $\epsilon(x,\Phi)\leq\varepsilon$ be large.

The two cardinal parameters in our problem are the sparsity factor \( \sparsity \) and the overcompleteness ratio \( \overcomp \), defined as
\begin{equation}\label{CardinalParametersDef}
\sparsity \triangleq \frac{k}{d}, \qquad \overcomp \triangleq \frac{n}{d}.
\end{equation}
The overcompleteness ratio can be thought of as the aspect ratio of the (wide) dictionary matrix $\Phi\in\R^{d\times n}$. Similarly, the sparsity factor $\sparsity$ is the aspect ratio of the (tall) sub-matrix of $\Phi$ containing the $k$ atoms participating in the decomposition of the signal $x$. Note that $\sparsity$ is not the percentage of nonzeros in the coefficient vector $\alpha$ in \eqref{RepresentationErrorEq} (which would be $k/n$).

Our goal is to characterize the minimal overcompletness $\overcomp$ such that all/most signals in $\R^d$ possess a sparse representation with sparsity $\sparsity$, up to some permissible error $\varepsilon$. Before we state our main results, let us first give some intuition into why this problem is not trivial.

\begin{figure}[!t]
	\centering
	\includegraphics[width=3.5in]{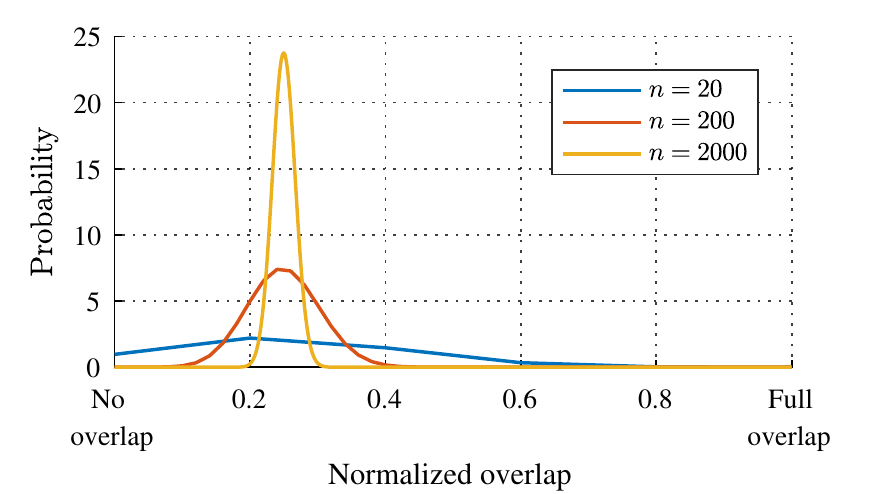}
	\caption{Given a set of $n$ atoms, there exist $\binom{n}{k}$ distinct subsets of $k$ atoms. The graph shows the relative amount of pairs of such subsets as a function of their overlap. As $n$ tends to infinity while keeping $k/n$ fixed, the relative overlap develops a sharp peak at $k/n$ and its standard deviation tends to $0$. Thus, when $n$ is large, practically all pairs of groups of $k$ atoms share $k\times (k/n)$ atoms. Here, $k/n=0.25$.}
	\label{fig:Hypergeometric}
\end{figure}

As mentioned above, each choice of $k$ atoms from the dictionary corresponds to a single subspace of dimension at most $k$. The volume of the set of signals whose normalized distance from this subspace is bounded by $\varepsilon$, can be computed in closed form (see Sec.~\ref{Derivations}). The problem is that when the number of atoms $n$ tends to infinity while keeping the ratio $\frac{k}{n}$ fixed, almost all pairs of groups of $k$ atoms from the dictionary share a significant number of atoms. That is, almost all pairs of subspaces intersect. These intersections cannot be disregarded, especially when seeking to upper-bound the minimal required $\overcomp$. Specifically, let $Q=\binom{n}{k}^2$ denote the total number of ordered pairs of subgroups of $k$ atoms and let $q(\ell)$ denote the number of pairs that share precisely $\ell$ atoms. Then, by definition, $q(\ell)/Q$ is a hypergeometric distribution with parameters $(n,k,k)$. It is well known that the mean of this distribution is $k\times \frac{k}{n}$ and that its standard deviation is $\sqrt{k}\times \sqrt{\frac{k}{n}(1-\frac{k}{n})\frac{n-k}{n-1}}$. Thus, normalizing by the maximal possible overlap $k$ and taking $n$ to infinity while keeping $\frac{k}{n}$ fixed, we obtain a probability distribution whose mean tends to $\frac{k}{n}$ and whose standard deviation tends to $0$ as $O(\frac{1}{\sqrt{k}})$. In other words, almost all pairs of groups of $k$ atoms share precisely $k\times \frac{k}{n}$ atoms in this regime. This phenomenon is illustrated in Fig.~\ref{fig:Hypergeometric}.

To overcome this difficulty, in our upper bounds analyses, we focus on a special type of (sub-optimal) structured dictionaries and also pose a certain (sub-optimal) restriction on the allowed choices of atoms from the dictionary. These assumptions significantly simplify the derivations, and while they may seem to lead to a crude overestimation of the required overcompleteness, we show that the resulting bounds are rather accurate in quite a wide range of settings.

\subsection{Worst-case analysis} \label{WorstCaseResults}
We begin by studying the problem from a worst-case standpoint.
\newtheorem{UniversalSparseRepresentation}[Definitions]{Definition}
\begin{UniversalSparseRepresentation}[Universal \(k\)-sparse representation dictionary] \label{UniversalSparseRepresentationDef}
	We say that \(\Phi \in \R^{d \times n}\) is a \emph{universal \(k\)-sparse representation dictionary with precision \(\varepsilon \)} if all signals in \(\R^d\) admit a \(k\)-sparse representation with precision \(\varepsilon\) over $\Phi$, namely \(\epsilon(x,\Phi)\leq\varepsilon,\, \forall x\in\R^d\) (equivalently, \(  \max_{x \in \R^d } \epsilon(x,\Phi)\leq\varepsilon  \)).
\end{UniversalSparseRepresentation}

Let us denote by \( \overcompD \) the minimal overcompleteness allowing universal sparse representation. In the theorems below we provide upper and lower bounds on $\overcompD$. These bounds are expressed only in terms of the sparsity $\sparsity$ and allowable error $\varepsilon$. Particularly, they are independent of the dimension \(d\) (despite the fact that $\overcompD$ itself does depend on $d$).

\newtheorem{WorstCaseLowerBound}[Theorems]{Theorem}
\begin{WorstCaseLowerBound}[Worst-case lower bound] \label{WorstCaseLowerBoundTheorem}
	If \( \varepsilon \in (0, \sqrt{1-\sparsity} ) \), then
	\begin{equation} \label{WorstCaseLowerBoundEq}
		\overcompD \geq c_1( \sparsity,\varepsilon ) \times \sparsity^{ \frac{3}{2} } \left(\frac{1}{\varepsilon}\right)^{ \frac{1}{\sparsity} - 1 },
	\end{equation}
	where $c_1( \sparsity,\varepsilon ) = e^{-1}\sqrt{(1-\sparsity)^{1/\sparsity-1}/({1-\varepsilon^2})} \geq e^{-\frac{3}{2}}$ for all \(s\) and \( \varepsilon \).
	If $\varepsilon \in (\sqrt{1-s},1) $, then
	\begin{equation}
		\overcompD = 1.
	\end{equation}
\end{WorstCaseLowerBound}

As expected, when either the sparsity factor $\sparsity$ or the precision $\varepsilon$ are small, the required overcompleteness is large. However, interestingly, the dependence on $\sparsity$ and $\varepsilon$ is quite different. While the bound is polynomial in $\varepsilon^{-1}$, it is exponential in $\sparsity^{-1}$, implying that universal sparse representation is practically impossible at very small sparsity factors.

\newtheorem{WorstCaseUpperBound}[Theorems]{Theorem}
\begin{WorstCaseUpperBound}[Worst-case upper bound] \label{WorstCaseUpperBoundTheorem}
	If \( \varepsilon \in (0, 1) \), \(k\) is a divisor of \(d\) and \( \sparsity \leq \frac{1}{3} \), then
	\begin{equation} \label{WorstCaseUpperBoundEq}
		\overcompD \leq c_2( \sparsity,\varepsilon ) \times \log (\sparsity^{-1}) \sparsity^{-\frac{1}{2}}  \left(\frac{1}{\varepsilon}\right)^{ \frac{1}{\sparsity} - 1 },
	\end{equation}
	where \(c_2( \sparsity,\varepsilon )= \sqrt{2\pi} (1 + \frac {2} {\log \sparsity^{-1}} ) (1 + \frac{\log \log \sparsity^{-1}}{\log \sparsity^{-1}} + \frac{\sqrt{e}} {\sparsity^{-1}} ) (1-\varepsilon^2 \frac{1-\sparsity}{1+\sparsity})^{\frac{1}{2}} \leq 12 \) for all \(s\) and \( \varepsilon \). If \(k\) is not a divisor of \(d\), then this bound holds true with \( \sparsity^{-1}  \) replaced by \( \ceil{\sparsity^{-1}}  \).
\end{WorstCaseUpperBound}

As can be seen, both bounds are exponentially equivalent\footnote{Namely, the ratio between the log of the bounds and the log of $ ( 1/\varepsilon ) ^{1/\sparsity-1} $ tends to 1 as either $\sparsity$ or $\varepsilon$ tend to zero.}  to $ ( 1/\varepsilon ) ^{1/\sparsity-1} $. This implies that under the conditions of Theorems~\ref{WorstCaseLowerBoundTheorem} and~\ref{WorstCaseUpperBoundTheorem}, the minimal overcompleteness $\overcompD$ satisfies
\begin{equation} \label{ExponentialEquality}
	\overcompD \approx \left(\frac{1}{\varepsilon}\right)^{\frac{1}{\sparsity} -1},
\end{equation}
where $\approx$ denotes exponential equivalence. One can think of the representation error as noise, in which case the term $1/\varepsilon$ can be interpreted as the signal-to-noise ratio (SNR). Therefore,~\eqref{ExponentialEquality} can also be written as
\begin{equation} \label{ExponentialEquality_SNR}
	\overcompD_{\text{dB}} \sim  \left( \frac{1}{\sparsity} -1 \right) \SNR_{\text{dB}} ,
\end{equation}
where \(\sim\) denotes asymptotic equivalence, $\SNR_{\text{dB}} = 20\log_{10}(1/\varepsilon)$ and $\overcompD_{\text{dB}} = 20\log_{10}(\overcompD)$.

Exponential equivalence is agnostic to polynomial dependencies. Thus, to refine our intuition, it is instructive to examine the ratio between the bounds \eqref{WorstCaseUpperBoundEq} and \eqref{WorstCaseLowerBoundEq}. As can be seen, this ratio is bounded from above as a function of  $\varepsilon$ and is only polynomial in $\sparsity^{-1}$ (behaves as \( \Theta( \log (\sparsity^{-1})  \sparsity^{-2}) \)). This indicates that our bounds are relatively accurate for moderate sparsity factors, even when $\varepsilon$ is small, but may become inaccurate for very small $\sparsity$.

\subsection{Average-case analysis} \label{AverageCaseResults}
The upper bound of Theorem \ref{WorstCaseUpperBoundTheorem} is rather pessimistic as it guarantees that \emph{all} signals can be sparsely represented with precision \(\varepsilon\), including esoteric and unlikely signals. In many practical situations, it may be enough to loosen this requirement and replace it by a probabilistic one. Specifically, suppose we have prior knowledge in the form of a distribution $\Omega$ over signals in $\R^d$. In this case, it may be enough to settle for dictionaries allowing sparse representation only \emph{with high probability}.
\newtheorem{OptimalSuccessProbability}[Definitions]{Definition}
\begin{OptimalSuccessProbability}[Optimal success probability] \label{OptSuccessProbDef}
	For any given distribution \( \Omega \) over \( \R^d \) and dictionary size \( d \times n \), we define the \emph{optimal success probability} under \(\Omega\) as
	\begin{equation}\label{OptSuccessProbEq}
	\optSuccProb{\Omega} \triangleq \max_{\Phi \in \R^{d \times n} } \prob{ \epsilon(x,\Phi) \leq \varepsilon },
	\end{equation}
	where \(x\) is a random vector with distribution $\Omega$.
\end{OptimalSuccessProbability}

To obtain bounds that do not depend on the prior $\Omega$, we will examine the worst-case optimal success probability over all possible distributions $\Omega$. Mathematically, let \( \mathcal{D} (\R ^d) \) be the collection of all distributions over \(\R ^d\). Then the worst-case optimal success probability is defined as
\begin{equation} \label{WorstCaseOptimalSuccessProb}
	\optSuccProbWC = \min_{\Omega \in \mathcal{D} (\R ^d) } \optSuccProb{\Omega}.
\end{equation}
Studying the behavior of $\optSuccProbWC$ is particularly interesting in high dimensions. In this setting there is a sharp transition between the regime of overcompleteness factors at which $\optSuccProbWC$ tends to $1$ and the regime at which it tends to $0$. We would therefore like to study the limit of \( \optSuccProbWC \) as \(d\) tends to infinity, while keeping the sparsity factor \( \sparsity \)  fixed. To this end, we denote by \( \overcompS \) the minimal overcompleteness such that \( \lim_{d \rightarrow \infty } \optSuccProbWC = 1 \).

There are several important distinctions between $\overcompS$ of the average case scenario and $\overcompD$ of the worst case setting. First, $\overcompS$ is only affected by typical signals, whereas $\overcompD$ takes into account all signals. Therefore, we necessarily have that $\overcompS\leq\overcompD$. Second, as can be seen from \eqref{OptSuccessProbEq}, it may be that for each distribution \( \Omega \) the optimal dictionary is different. Thus, as opposed to the worst-case analysis, in the average case setting we do not guarantee the existence of a single dictionary that is good for all signals. Finally, $\overcompS$ is defined only for $d\rightarrow\infty$, whereas $\overcompD$ is defined for all $d$. This is particularly important when bounding these quantities from above, since the minimal required overcompleteness becomes smaller as $d$ increases. This further contributes to our ability to obtain an upper bound on $\overcompS$, which is lower than the upper bound on $\overcompD$ in Theorem~\ref{WorstCaseUpperBoundTheorem}.

The next two theorems are analogous to Theorems~\ref{WorstCaseLowerBoundTheorem} and~\ref{WorstCaseUpperBoundTheorem}. The first statement in each theorem bounds \( \overcompS \), and thus provides an asymptotic analysis. The second statement characterizes the convergence to the asymptotic behavior, and is relevant for any finite dimension.
\newtheorem{AverageCaseLowerBound}[Theorems]{Theorem}
\begin{AverageCaseLowerBound}[Average-case lower bound] \label{AverageCaseLowerBoundTheorem}
	If \( \varepsilon \in (0, \sqrt{1-\sparsity} ) \), then
	\begin{equation} \label{AverageCaseLowerBoundEq}
		\overcompS \geq c_1( \sparsity,\varepsilon ) \times \sparsity^{ \frac{3}{2} } \left(\frac{1}{\varepsilon}\right)^{ \frac{1}{\sparsity} - 1 },
	\end{equation}
	where \( c_1( \sparsity,\varepsilon )\) is as in \eqref{WorstCaseLowerBoundEq}.	Furthermore, for any finite dimension \( d \),
	\begin{equation} \label{AverageCaseLowerBoundProb}
	\optSuccProbWC  \leq \frac{1}{\sqrt{2\pi \sparsity \left(1 - \frac{\sparsity}{\overcomp}\right)}} \times  d^{-\frac{1}{2}}\exp\{ - c_3( \sparsity,\varepsilon, \overcomp ) d \} .
	\end{equation}
	Here, \( c_3( \sparsity,\varepsilon, \overcomp ) = \frac{1}{2} \relent{1-s}{\varepsilon^2} - \overcomp \entropyi{ \frac{\sparsity}{\overcomp} } \), where $\entropyi{\alpha} =-\alpha\log(\alpha) -(1-\alpha)\log(1-\alpha)$ is the entropy of a $\text{Bernoulli}(\alpha)$ random variable, and \( \relent{\alpha}{\beta} = \alpha\log(\frac{\alpha}{\beta}) + (1-\alpha)\log(\frac{1-\alpha}{1-\beta}) \) is the Kullback Leibler divergence between the \( \text{Bernoulli}(\alpha)\) and \(\text{Bernoulli}(\beta)\) distributions.
\end{AverageCaseLowerBound}

In fact, as we show in Sec.~\ref{AverageCaseLowerBoundAnalysis}, when the overcompleteness \( \overcomp \) is smaller than the right hand side of \eqref{AverageCaseLowerBoundEq}, not only that $\optSuccProbWC$ does not tend to $1$, it actually tends to $0$. This implies that below this bound, universal sparse representation is practically impossible in high dimensions (for the worst case distribution).

The next theorem is stated in terms of the incomplete regularized beta function $\Ii{x}{\alpha}{\beta}$, which is the probability that a \( \myBetai{\alpha}{\beta}\) random variable is smaller than $x$.
\newtheorem{AverageCaseUpperBound}[Theorems]{Theorem}
\begin{AverageCaseUpperBound}[Average-case upper bound] \label{AverageCaseUpperBoundTheorem}
	If \( \sparsity = \frac{1}{m} \) for some \( m \in \N \), then
	\begin{equation} \label{AverageCaseUpperBoundEq}
		\overcompS \leq c_4( \sparsity,\varepsilon ) \times \sparsity^{\frac{1}{2}}  \left(\frac{1}{\varepsilon}\right)^{ \frac{1} {\sparsity} - 1 },
	\end{equation}
	where  \(c_4( \sparsity,\varepsilon )= \sqrt{\tfrac{\pi}{2} (1-\varepsilon^2 (1-\sparsity)/(1+\sparsity) ) }\leq \sqrt{\tfrac{\pi}{2}} \)  for all \(s\) and \( \varepsilon \).
	Furthermore, for any finite dimension \( d \), if \( \varepsilon^2 \leq \tfrac{1}{2} \) and the overcompleteness ratio satisfies
	\begin{equation} \label{AverageCaseUpperBoundProb1}
		\overcomp \geq \frac{\sparsity} { \I{\delta^2}{\frac{1}{2}(\frac{1}{s}-1)}{ \frac{1}{2} } } = \bigtheta{  \sparsity^{\frac{1}{2}}  \left(\frac{1}{\delta}\right)^{ \frac{1} {\sparsity} - 1 } }
	\end{equation}
	for some \( \delta \in (0, \varepsilon ) \), then
	\begin{equation} \label{AverageCaseUpperBoundProb2}
		\optSuccProbWC \geq 1 - \frac{c_5(\sparsity,\varepsilon,\overcomp,\delta)}{c_5(\sparsity,\varepsilon,\overcomp,\delta) + d},
	\end{equation}
	where \(  c_5(\sparsity,\varepsilon,\overcomp,\delta) =   \frac{ (1-2\varepsilon^2) (\Ii{1-2\varepsilon^2} {\frac{1}{2}}{\frac{1-\sparsity}{2\sparsity}})^{ \frac{\overcomp}{ \sparsity } } + \varepsilon^4} {(\varepsilon^2-\delta^2)^2} \frac{(1+2\sparsity)}{\sparsity } \) is a constant independent of the dimension \(d\).
\end{AverageCaseUpperBound}

Note that Theorem \ref{AverageCaseLowerBoundTheorem} provides a \emph{lower bound} on the minimal required overcompleteness (see \eqref{AverageCaseLowerBoundEq}) and an \emph{upper bound} on the probability of success (see \eqref{AverageCaseLowerBoundProb}). Similarly, Theorem \ref{AverageCaseUpperBoundTheorem} provides an \emph{upper bound} on the minimal required overcompletness (see \eqref{AverageCaseUpperBoundEq}), which is further refined via a \emph{lower bound} on the probability of success (see \eqref{AverageCaseUpperBoundProb1},\eqref{AverageCaseUpperBoundProb2}).

Comparing Theorems \ref{WorstCaseUpperBoundTheorem} and \ref{AverageCaseUpperBoundTheorem}, it can be seen that the average-case analysis provides an improvement of \( \Theta (\sparsity^{-1} \log \sparsity^{-1} ) \) over the worst-case analysis (compare \eqref{WorstCaseUpperBoundEq} with \eqref{AverageCaseUpperBoundEq}). Another interesting comparison is between the lower and upper bounds in the average-case scenario. In general, the ratio between \eqref{AverageCaseLowerBoundEq} and \eqref{AverageCaseUpperBoundEq} is a complicated function of \( \sparsity \) and \( \varepsilon \) which behaves as \( \Theta( \sparsity^{-1} ) \). Yet, for certain regimes of \( \varepsilon \) and \( \sparsity \) we can obtain simple expressions. When \( \varepsilon \ll 1 \), the ratio becomes approximately \( \sqrt{ \frac{\pi e^2 }{2(1-\sparsity)^{\sparsity^{-1}-1}  } } \sparsity^{-1} \). This expression is independent of \( \varepsilon \), implying that the bounds are relatively tight for moderate values of \( \sparsity \) when \( \varepsilon \) is small. Similarly, when \( \sparsity \ll 1 \), the ratio becomes approximately \( \sqrt{ \frac{\pi e^3}{2} } (1-\varepsilon^2) \sparsity^{-1} \).

\section{Derivations} \label{Derivations}
In this section, we prove Theorems~\ref{WorstCaseLowerBoundTheorem}-\ref{AverageCaseUpperBoundTheorem}. As discussed above, each choice of $k$ atoms from a dictionary \( \Phi \in \R ^{d \times n } \), allows to perfectly represent all signals lying in some linear subspace of dimension at most $k$. Therefore, the set of all signals admitting a $k$-sparse representation over $\Phi$ corresponds to the union of these \( N =  \binom{n}{k} \) linear subspaces, which we denote by \( \{ \psi_i\}_{i=1}^N \). In our setting, we allow for \emph{approximate} representations with a relative error of at most $\varepsilon$. To this end, we use the notion of spherical dilations. Specifically, the spherical dilation of a set $X\subset \R^d$ with radius \(r\) is defined as the set
\begin{equation} \label{BallDef}
	\ball{X}{r} = \{ y \in \R^d : \exists x \in X \quad \text{s.t.} \quad \norm{ y - x } \leq r \}.
\end{equation}
In order for \( \Phi \) to be a universal \( k \)-sparse representation dictionary with precision \( \varepsilon \), each point on the unit sphere must be contained in at least one of the \( \varepsilon \)-dilations of its subspaces \( \{ \psi_i\}_{i=1}^N \). In other words, the universal sparse representation problem is in fact a covering problem as we are interested in covering the unit sphere by \( \varepsilon \)-dilations of linear subspaces. That is, we would like that
\begin{equation} \label{SphereCover}
	S^{d-1} \subseteq \bigcup\limits_{i=1}^{N} \ball{\psi_i}{\varepsilon},
\end{equation}
where $S^{d-1}$ is the unit sphere in $\R^d$.

\begin{figure}
	\centering
	\includegraphics[trim=32 47 23 40, clip, width=3.5in]{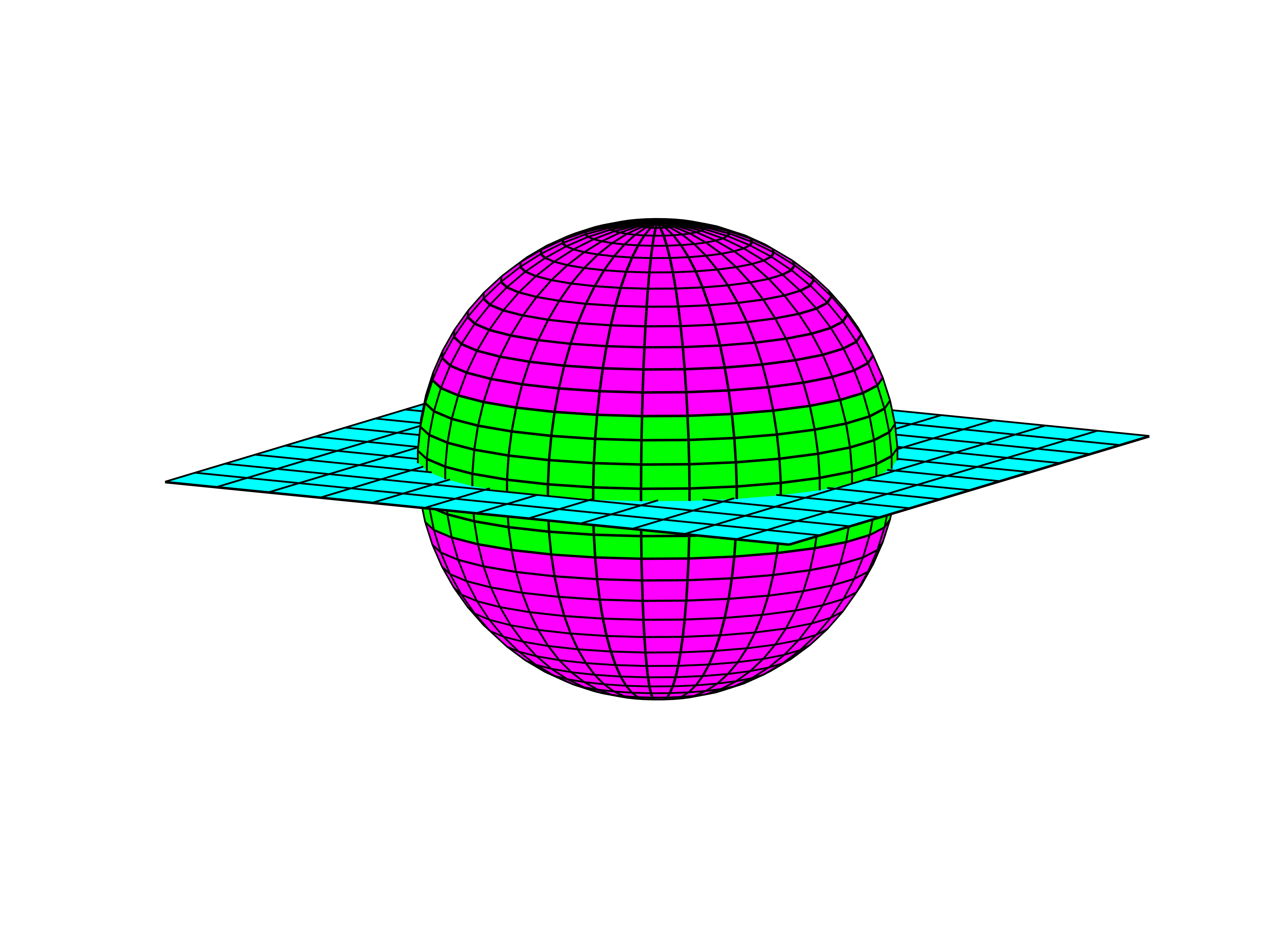}
	\caption{The area from the unit sphere in $\R^3$ covered by an $\varepsilon$-dilation of a 2D subspace is shown in green. Corollary~\ref{SubspaceCoverageCorollary} gives an explicit formula for the ratio between the green area and the surface area of the unit sphere, for any ambient dimension $d$, subspace dimension $k$, and dilation $\varepsilon$.}\label{fig:ball}
\end{figure}

Let us start by examining the relative area covered by an \( \varepsilon \)-dilation of a single linear subspace\footnote{A similar analysis was provided in \cite{akccakaya2008frame} for complex spaces.} $\psi$ (namely, the ratio between the area of $\ball{\psi}{\varepsilon}\cap S^{d-1}$ and the area of $S^{d-1}$). This relative area, illustrated in Fig.~\ref{fig:ball}, can be interpreted as the probability that a point \(x\) chosen uniformly at random from the unit sphere $S^{d-1}$ be \( \varepsilon \)-close to \( \psi \). The square distance between \( x \) and \( \psi \) is \(\normi{x-\proj{x}{\psi}}^2=\normi{x}^2 - \normi{\proj{x}{\psi}}^2\), where \(\proj{}{\psi}\) is the orthogonal projection matrix onto $\psi$. Thus, since \(\normi{x} = 1\), \(x\) is \( \varepsilon \) close to \( \psi \) if and only if \( \normi{\proj{x}{\psi}}^2 \geq 1-\varepsilon^2 \). This implies that we have the relation
\begin{equation} \label{FundamentalEquality}
	\frac{\area{\ball{\psi}{\varepsilon} \cap S^{d-1}}}{\area{S^{d-1}}} = \prob{\norm{\proj{x}{\psi}}^2 \geq 1-\varepsilon^2},
\end{equation}
where $\area{V}$ denotes the area of the $d-1$ dimensional manifold $V\subset \R^d$. Our problem thus boils down to determining the distribution of the length of a random vector with uniform distribution on the unit sphere, projected down onto a fixed \(k\)-dimensional subspace. This result can be found \eg in \cite{muirhead1982aspects}. For our purposes, it is more convenient to state the result for the equivalent setting where the subspace is random and the unit vector is deterministic.
\newtheorem{RandomSubspaceProjection}[Lemmas]{Lemma}
\begin{RandomSubspaceProjection}[Random subspace projection \cite{muirhead1982aspects}] \label{RandomSubspaceProjectionTheorem}
	Let \( \mathcal{V} \) denote the \( k \)-dimensional subspace spanned by a set \( \{\nu_i\}_{i=1}^k \) of independent isotropically distributed random vectors in \( \R^d \). Then for every deterministic unit-norm vector \(x \in S^{d-1}\),
	\begin{equation} \label{RandomSubspaceProjectionDistribution}
		\norm{\proj{x}{  \mathcal{V} }}^2 \thicksim \myBeta{\frac{k}{2}}{\frac{d-k}{2}}.
	\end{equation}
\end{RandomSubspaceProjection}

Note that the distribution of \( \normi{\proj{x}{ \mathcal{V}}}^2 \) is independent of \( x \), hence this result also holds true when $x$ is a random unit norm vector that is independent of \(\{\nu_i\}\).

From \eqref{RandomSubspaceProjectionDistribution} we have that \( \probi{\normi{\proj{x}{\psi}}^2 \geq 1-\varepsilon^2} = 1-\Ii{1-\varepsilon^2}{\frac{k}{2}}{\frac{d-k}{2}} \), where $\Ii{x}{\alpha}{\beta}$ is the cumulative distribution function of the $\myBetai{\alpha}{\beta}$ distribution, also known as the regularized incomplete beta function. From the properties of the beta distribution, we can also write \( 1-\Ii{1-\varepsilon^2}{\frac{k}{2}}{\frac{d-k}{2}} = \Ii{\varepsilon^2} {\frac{d-k}{2}}{\frac{k}{2} }\). Thus, from \eqref{FundamentalEquality} and \eqref{RandomSubspaceProjectionDistribution} we reach the following conclusion.
\newtheorem{SubspaceCoverage}[Corollaries]{Corollary}
\begin{SubspaceCoverage}[Subspace coverage] \label{SubspaceCoverageCorollary}
	Let \( \psi \) be a \( k \) dimensional linear subspace in \( \R^d \). Then the relative area covered by \( \ball{\psi}{\varepsilon} \) from the unit sphere is precisely \( \Ii{\varepsilon^2 }{\frac{d-k}{2}}{\frac{k}{2}} \).
\end{SubspaceCoverage}

We remark that this corollary can be seen as a generalization of \cite{li2011concise}, which proved it for one dimensional subspaces.

We are now ready to prove the theorems of Section \ref{MainResults}. In sections \ref{WorstCaseLowerBoundAnalysis} and \ref{WorstCaseUpperBoundAnalysis} we prove the lower and upper bounds, respectively, for the worst-case setting. Similarly, in sections \ref{AverageCaseLowerBoundAnalysis} and \ref{AverageCaseUpperBoundAnalysis} we derive the necessary and sufficient conditions, respectively, for the average-case setting.

\subsection{Worst-case lower bound}\label{WorstCaseLowerBoundAnalysis}
In this section we prove Theorem \ref{WorstCaseLowerBoundTheorem}. Following the discussion above, it is clear that \( \Phi \) is \emph{not} a universal \(k\)-sparse representation dictionary if the unit sphere \( S^{d-1} \) is not contained in the union of the \( \varepsilon \) dilations of its $\binom{n}{k}$ subspaces $\{\psi_i\}$, namely
\begin{equation}\label{WorstCaseLowerBoundCondition}
	S^{d-1} \not\subset \bigcup\limits_{i=1}^{N} \ball{\psi_i}{\varepsilon}.
\end{equation}
This condition is satisfied when
\begin{equation}\label{WorstCaseLowerBoundCondition2}
	\frac{\area{\bigcup\limits_{i=1}^{N} \ball{\psi_i}{\varepsilon}  \cap S^{d-1}}}{\area{S^{d-1}}}  < 1.
\end{equation}
Applying the union bound on the left hand side of \eqref{WorstCaseLowerBoundCondition2} gives
\begin{equation}\label{WorstCaseUnionBound}
	\frac{\area{\bigcup\limits_{i=1}^{N} \ball{\psi_i}{\varepsilon}  \cap S^{d-1}}}{\area{S^{d-1}}} \leq \sum\limits_{i=1}^{N} \frac{\area{ \ball{\psi_i}{\varepsilon}  \cap S^{d-1}}}{\area{S^{d-1}}}.
\end{equation}
From Corollary \ref{SubspaceCoverageCorollary}, each of the summands in the right hand side of \eqref{WorstCaseUnionBound} is equal to \( \Ii{\varepsilon^2 }{\frac{d-k}{2}}{\frac{k}{2}} \). Therefore,  a sufficient condition for \eqref{WorstCaseLowerBoundCondition2} to hold, is that
\begin{equation}\label{WorstCaseLowerBoundCondition3}
	\binom{n}{k} \,\I{\varepsilon^2}{\frac{d-k}{2}} {\frac{k}{2}} < 1.
\end{equation}
To simplify this expression, let us further bound the terms \( \binom{n}{k}  \) and \( \Ii{\varepsilon^2 } {\frac{d-k}{2}} {\frac{k}{2}} \). For the binomial coefficient, we use the bound
\begin{equation}\label{BinomialUpperBound}
	\binom{n}{k} < \left( \frac{ne}{k} \right)^k = \exp\left\{k \left(\log\left(\frac{n}{k}\right)+1\right)\right\}.
\end{equation}
For the incomplete beta function, we use the following tail bound from \cite{dasgupta2003elementary}.
\newtheorem{IncomBetaUpperBound}[Lemmas]{Lemma}
\begin{IncomBetaUpperBound}[\cite{dasgupta2003elementary}]\label{IncomBetaUpperBoundTheorem}
	 Let \( L \thicksim \myBetai{\frac{k}{2}}{\frac{d-k}{2}} \), where \(k<d\) are natural numbers. Then for every \( \beta > 1 \),
	 \begin{equation}
		 \prob{L \geq \frac{\beta k}{d}} \leq \beta^{ \frac{k}{2} } \left( 1+ \frac{(1-\beta)k}{d-k} \right)^{\frac{d-k}{2}}.
	 \end{equation}
\end{IncomBetaUpperBound}
To match our setting in \eqref{FundamentalEquality}, we choose \( \frac{\beta k}{d} = 1-\varepsilon^2  \). This relation implies that $\beta = \frac{1-\varepsilon^2}{\sparsity} $  and \( 1+ \frac{(1-\beta)k}{d-k} = \frac{\varepsilon^2}{1-\sparsity} \). Note that the lemma applies to $\beta>1$, which translates to the requirement that \( \sparsity < 1- \varepsilon^2 \). Applying the lemma, gives us the bound
\begin{equation} \label{BetaUpperBound}
	\I{\varepsilon^2 }{\frac{d-k}{2}}{\frac{k}{2}} \leq \left( \frac{1-\varepsilon^2}{\sparsity} \right)^{ \frac{k}{2} } \left( \frac{\varepsilon^2}{1-\sparsity} \right)^{\frac{d-k}{2}}.
\end{equation}
To present this bound more compactly, we use the function \( \relent{\alpha}{\beta} = \alpha\log(\frac{\alpha}{\beta}) + (1-\alpha)\log(\frac{1-\alpha}{1-\beta}) \), which is the Kullback Leibler divergence between the \( \text{Bernoulli}(\alpha)\) and \(\text{Bernoulli}(\beta)\) distributions. Then, the right-hand-side of \eqref{BetaUpperBound} can be equivalently expressed as
\begin{equation}\label{BetaUpperBound2}
	\left( \frac{1-\varepsilon^2}{\sparsity} \right)^{ \frac{k}{2} } \left( \frac{\varepsilon^2}{1-\sparsity} \right)^{\frac{d-k}{2}}  = \exp\left\{-\frac{d}{2} \relent{1- \sparsity }{\varepsilon^2}\right\}.
\end{equation}
Using the bounds \eqref{BinomialUpperBound} and \eqref{BetaUpperBound},\eqref{BetaUpperBound2} in \eqref{WorstCaseLowerBoundCondition3}, we conclude that the relative area is upper-bounded by
\begin{equation}\label{WorstCaseLowerBoundCondition4}
\begin{aligned}
	\exp\left\{ -\frac{d}{2} \relent{1-s}{\varepsilon^2}+k\left(\log\left(\frac{n}{k}\right)+1\right) \right\}.
\end{aligned}
\end{equation}
Note from \eqref{CardinalParametersDef} that \(\frac{n}{k}=\frac{\overcomp}{\sparsity} \). Consequently, the argument of the exponent in \eqref{WorstCaseLowerBoundCondition4} can be written as
\begin{align}
	-k\left(\frac{1}{2\sparsity} \relent{1-\sparsity}{\varepsilon^2} - \log\left(\frac{\overcomp}{\sparsity}\right)-1 \right).
\end{align}
Therefore, to guarantee that the relative area is strictly smaller than $1$, it is sufficient to require that \( \frac{1}{2\sparsity} \relent{1-\sparsity}{\varepsilon^2} - 1 \geq \log(\frac{\overcomp}{\sparsity}) \), implying that if
\begin{equation}\label{eq:WorstCaseLowerBoundExp}
	\overcomp \leq \sparsity \exp\left\{\frac{1}{2\sparsity}\relent{1-\sparsity}{\varepsilon^2}-1\right\},
\end{equation}
then universal sparse representation is impossible. Substituting the expression for $\relent{1-\sparsity}{\varepsilon^2}$ in \eqref{eq:WorstCaseLowerBoundExp} gives the right-hand side of \eqref{WorstCaseLowerBoundEq}, thus completing the proof of the first part of Theorem~\ref{WorstCaseLowerBoundTheorem}. Note that the constant $c_1(\sparsity,\varepsilon)$ in \eqref{WorstCaseLowerBoundEq} satisfies
\begin{align}
c_1(\sparsity,\varepsilon) \geq c_1(\sparsity,0) \geq \lim_{\sparsity\to 0} c_1(\sparsity,0) = e^{-\frac{3}{2}}
\end{align}
for every $\varepsilon$ and $\sparsity$ in the range $(0,1)$.

We next examine the case \( \varepsilon^2>1-\sparsity \). On one hand, if the number of atoms \( n \) is smaller than the dimension \( d \), then there exists a linear subspace in \( \R^d \) which is orthogonal to all atoms. Signals from this subspace will have a representation error of~1. Therefore, we must have that \( \overcompD \geq 1 \). On the other hand, we will show that a dictionary with \( d \) atoms is sufficient for representing all signals in \( \R^d \) with precision \( \sqrt{1-\sparsity} \). Specifically, take the dictionary to be the \( d \) dimensional identity matrix \( I_{d  \times d } \). It is easy to see that in this setting, the representation error is
\begin{equation} \label{eq:ErrorForIdentityMatrix}
\epsilon^2\left(x , I_{d  \times d } \right) = \min_{ |\kappa| = d-k } \sum_{i \in \kappa  } x^2_{ i }.
\end{equation}
Therefore, the worst case error can be written as
\begin{equation} \label{eq:WorstCaseSignal}
\max_{x  \in S^{d-1} } \epsilon^2\left(x , I_{d  \times d } \right)
= 1- \min_{x  \in S^{d-1} } \max_{ |\kappa| = k } \sum_{i \in \kappa  } x^2_{ i },
\end{equation}
where we used the fact that \( \norm{x}^2 = 1 \). Denoting \( \xi_i = x^2_i  \), the right hand term reduces to
\begin{equation} \label{eq:ConvexOptimization}
\min_{\xi \in \Delta^d } \max_{ |\kappa| = k } \sum_{i \in \kappa  } \xi_i,
\end{equation}
where \( \Delta^d \) is the \( d \) dimensional unit simplex. It is well known that max-\(k\)-sums are convex functions and that the unit simplex is a convex set. Therefore, this is a convex optimization problem. Due to the fact that max-\(k\)-sums are invariant under permutation of the components, this optimization function has a minimizer \( \xi^* \) that satisfies
\begin{equation}
\xi^*_i = \xi_0
\end{equation}
for all $ i \in\{1,\ldots,d\} $, for some constant \(\xi_0\). From the constraints, we have that \( \xi_0 = \tfrac{1}{d} \). Substituting this solution in \eqref{eq:WorstCaseSignal} we obtain
\begin{equation}
\max_{x  \in S^{d-1} } \epsilon^2\left(x , I_{d  \times d } \right) = 1-\frac{k}{d}=1-\sparsity,
\end{equation}
which is less than \( \varepsilon^2 \) by assumption. Therefore, the required overcompleteness \( \overcompD \) must be equal to 1.

\subsection{Worst-case upper bound}\label{WorstCaseUpperBoundAnalysis}
Next we prove Theorem \ref{WorstCaseUpperBoundTheorem}. As discussed above, a sufficient condition for universal sparse representation corresponds to covering the unit sphere \( S^{d-1} \) (see \eqref{SphereCover}). To obtain an upper bound, we will restrict attention to suboptimal dictionaries with a block-diagonal structure
\begin{equation}\label{MyDictionary}
	\Phi =
	\begin{bmatrix}
		\Phi_{1} & 0 & \dots  & 0 \\
		0 & \Phi_{2} & \dots  & 0 \\
		\vdots  & \vdots & \ddots & \vdots \\
		0 & 0 & \dots  & \Phi_{k}
	\end{bmatrix},
\end{equation}
where \( \frac{d}{k} \) is a natural number and each $\Phi_i$ is a $\frac{d}{k} \times \frac{n}{k}$ matrix with columns $\{\varphi^{(i)}_j\}_{j=1}^{\frac{n}{k}}$. Interestingly, it turns out that the degradation in performance incurred by using such dictionaries is moderate, as we show experimentally in Sec.~\ref{AverageCaseUpperBoundAnalysis}. Let us denote the corresponding partitioning of the signal \( x \) as
\begin{equation}\label{MySignal}
	x^T = [x_1^T \quad x_2^T \quad \cdots \quad x_k^T],
\end{equation}
where each $x_i$ is of length $\frac{d}{k}$.

The block-diagonal structure \eqref{MyDictionary} allows us to simplify the problem of covering by dilations of subspaces into a problem of covering by balls. Specifically, \(\Phi\) of~\eqref{MyDictionary} is a universal \(k\) sparse representation dictionary with precision \(\varepsilon \) for signals in $\R^d$ if the atoms of each sub-dictionary \(\Phi_i\) form an \(\varepsilon\) ball covering of the unit sphere in \(\R^{ \frac{d}{k}}\). Indeed, if each \(\Phi_i\) forms an \(\varepsilon\) ball covering, then it contains at least one atom whose relative error from \(x_i\) is less than \(\varepsilon \). Choosing this single atom from each dictionary, we obtain a $k$-sparse representation with a squared relative error of
\begin{align} \label{ModificationInequality}
\frac{\|x-\Phi\alpha\|^2}{\|x\|^2} &= \frac{\sum_{i=1}^{k}\limits \min_{1 \leq j \leq m }\limits \norm{x_i - \proj{x_i}{\varphi^{(i)}_j}}^2}{\norm{x}^2 } \nonumber\\
&\leq \frac{\sum_{i=1}^{k}\limits\norm{x_i}^2\varepsilon^2}{\|x\|^2} = \varepsilon^2.
\end{align}
Hence, our problem is reduced to finding the minimal number of balls \(m=\frac{n}{d}\) with radius \( \varepsilon\) that suffice to cover the unit sphere in \(\R^{\frac{d}{k}} \).

A lot of work has been done in the field of ball covering. In particular, several papers have derived bounds on the minimal covering \emph{density} \(\nu \) \cite{coxeter1959covering,rogers1963covering, boroczky2003covering, dumer2007covering,boroczky2004finite}. In our terminology, the covering density is the sum of all relative surface areas from the unit sphere that are covered by the balls. These \(m\) relative surface areas are all the same, and according to Corollary~\ref{SubspaceCoverageCorollary} are given by\footnote{Note that the area covered by a subspace of dimension $1$ spanned by a vector $\psi$, corresponds to all unit vectors that are $\varepsilon$ close to either $\psi$ or $-\psi$, hence the factor $\frac{1}{2}$.} \( \frac{1}{2} \Ii{\varepsilon^2 } {\frac{1-\sparsity}{2\sparsity}}{\frac{1}{2}}  \). Therefore, \( \nu = \frac{m}{2}\Ii{\varepsilon^2} {\frac{1-\sparsity}{2\sparsity}} {\frac{1}{2}} \). In \cite{boroczky2003covering} it was proven that for all \( \sparsity^{-1}=\frac{d}{k} \geq 3 \), the minimal covering density for this problem is bounded by
\begin{equation}
	\nu \leq h(\sparsity^{-1})\log(\sparsity^{-1})\sparsity^{-1},
\end{equation}
where \( h(x)= (1 + \frac {2} {\log x} ) (1 + \frac{\log \log x}{\log x} + 	\frac{\sqrt{e}} {x\log x} ) \). From this bound we get that
\begin{equation}
m = \frac{2h(\sparsity^{-1})\log(\sparsity^{-1})\sparsity^{-1}}{\I{\varepsilon^2 } { \frac{1-\sparsity}{2\sparsity}}{\frac{1}{2} }}
\end{equation}
atoms per sub-dictionary suffice. Since $n=mk$, the corresponding overcompleteness is $\overcomp=n/d=mk/d=m\sparsity$. We thus showed that if
\begin{equation}\label{eq:WorstCaseUpperBoundBeta}
\overcomp \geq \frac{2 h(\sparsity^{-1})\log(\sparsity^{-1})}{\I{\varepsilon^2 }  {\frac{1-\sparsity}{2\sparsity}} {\frac{1}{2}} },
\end{equation}
then universal sparse representation is possible.

The bound~\eqref{eq:WorstCaseUpperBoundBeta} can be easily computed for any $\varepsilon$ and $\sparsity$. However, its asymptotic behavior for small $\varepsilon$ and $\sparsity$ is hard to interpret. To obtain a simpler expression, let us use the following lemma (see proof in Appendix \ref{BetaBoundProof}).
\newtheorem{BetaBoundPF}[Lemmas]{Lemma}
\begin{BetaBoundPF}[Lower bound on incomplete beta function] \label{BetaLowerBoundLemma}
	Let \( a >0\), \( 0 < b \leq 1 \) and \(x \in [0,1] \), then
	\begin{equation} \label{BetaLowerBound}
	\Ii{x}{a}{b} \geq \frac{x^a}{ \Gamma(b) \left( (a+b)( 1 - x \tfrac{a}{a+1} ) \right)^{1-b} }.
	\end{equation}
\end{BetaBoundPF}

By setting \( a = \frac{1-\sparsity}{2\sparsity} \), \( b = \frac{1}{2} \) and \( x = \varepsilon^2\) we get
\begin{equation}
\I{\varepsilon^2} {\frac{1-\sparsity}{2\sparsity}} {\frac{1}{2}} \geq \sqrt{\frac{2\sparsity}{\pi(1-\varepsilon^2\frac{1-\sparsity}{1+\sparsity})}  } \varepsilon^{\sparsity^{-1}-1},
\end{equation}
where we used the fact that $\Gamma(\frac{1}{2})=\sqrt{\pi}$. Therefore, we have
\begin{equation}
	\overcomp \geq \sqrt{2\pi(1-\varepsilon^2 \tfrac{1-\sparsity}{1+\sparsity})} h(\sparsity^{-1}) \log (\sparsity^{-1}) \sparsity^{-\frac{1}{2}}  \left(\frac{1}{\varepsilon}\right)^{ \sparsity^{-1} - 1 },
\end{equation}
which demonstrates \eqref{WorstCaseUpperBoundEq} and thus completes the proof of Theorem~\ref{WorstCaseUpperBoundTheorem}. It is easily verified that $c_2(\sparsity,\varepsilon)$ of \eqref{WorstCaseUpperBoundEq} satisfies $c_2(\sparsity,\varepsilon)\leq c_2(\frac{1}{3},0) < 12$ for every $\varepsilon\in(0,1)$ and $\sparsity\in(0,\frac{1}{3})$.

A few words are in place regarding the tightness of the tail bound in Lemma \ref{BetaLowerBoundLemma}, at least for the case $b=\frac{1}{2}$. From Corollary~\ref{SubspaceCoverageCorollary}, we have that the relative surface area covered from the unit sphere in \( \R^d \) by a spherical cap with half chord~\( \varepsilon \), is given by \( \frac{1}{2} \Ii{\varepsilon^2}{\frac{d-1}{2}}{\frac{1}{2}} \). This geometric quantity arises in many fields, including machine learning \cite{hanneke2014theory,safran2016quality}, estimation theory \cite{ramirez2012low}, communication \cite{chaaban2016free}, and even systematic biology~\cite{klingenberg2013evolutionary}. As a result, obtaining tight bounds for this quantity, has attracted interest by various researchers, \eg \cite[Corollary 3.2]{boroczky2003covering}, \cite[Th. 3.1]{frankl1990some}. To the best of our knowledge, the tightest bounds are found in \cite{frankl1990some}, where it has been shown that for all \( \varepsilon \in (0,1) \) and \( d \in \N \), this relative area is lower bounded by
\begin{align} \label{eq:FranklBounds}
	\frac{1}{2} \I{\varepsilon^2}{\frac{d-1}{2}}{\frac{1}{2}}> \frac{  1 - \left( 1- \frac{1}{\sqrt{d}} \right)^{ \frac{d-1}{2} } }{\sqrt{2 \pi \frac{(d-1)^2}{d-2}\left(1-\varepsilon^2\left(1- \frac{1}{\sqrt{d}}\right)\right)}} \varepsilon^{d-1}.
\end{align}
Our lemma provides the lower bound
\begin{equation}\label{eq:OurLowerBoundBeta}
	\frac{1}{2} \I{\varepsilon^2}{\frac{d-1}{2}}{\frac{1}{2}}\geq\frac{\varepsilon^{d-1} }{ \sqrt{ 2\pi d (1-\varepsilon^2\frac{d-1}{d+1}) }} .
\end{equation}
It is easily verified that \( \frac{(d-1)^2}{d-2}> d \), \( 1-d^{- \frac{1}{2} } \leq \frac{d-1}{d+1}\) and \(  ( 1- d^{- \frac{1}{2} } )^{ \frac{d-1}{2} }>0 \). Thus, our lower bound is higher than \eqref{eq:FranklBounds} for all \( \varepsilon \) and \( d \). The authors of \cite{frankl1990some} also derived the upper bound
\begin{equation} \label{eq:FranklBounds2}
		\frac{1}{2} \I{\varepsilon^2}{\frac{d-1}{2}}{\frac{1}{2}} < 		
		\frac{\varepsilon^{d-1}}{\sqrt{2\pi(d-1)(1-\varepsilon^2)}}.
\end{equation}
It can be seen that for high dimensions, the ratio between our lower bound and this upper bound tends to one, implying that they are asymptotically tight (the tightness of \eqref{eq:FranklBounds2} was also noted in \cite{frankl1990some}).

\subsection{Average-case lower bound} \label{AverageCaseLowerBoundAnalysis}
We now turn to prove Theorem \ref{AverageCaseLowerBoundTheorem}, which provides an average-case lower bound on the overcompleteness for the setting in which $x$ is a random vector. As mentioned in Section~\ref{MainResults}, to obtain bounds that do not depend on the distribution of $x$, we consider the worst case distribution (\textit{i.e.} the one leading to the lowest optimal success probability). We begin with the following observation, whose proof is provided in Appendix~\ref{WorstCaseDistributionProof}.
\newtheorem{WorstCaseDistribution}[Lemmas]{Lemma}
\begin{WorstCaseDistribution}[Worst case distribution] \label{WorstCaseDistributionLemma}
	Any isotropic distribution \( \Theta\) achieves the \emph{minimum} of the optimal success probability, \textit{i.e.}
	\begin{equation}
		\optSuccProbWC = \optSuccProb{\Theta}.
	\end{equation}
\end{WorstCaseDistribution}

This lemma shows that we can safely focus on the case where \( x \thicksim N(0,I_{d \times d})\), which is an isotropic distribution. Our derivation is similar to the one in Sec.~\ref{WorstCaseLowerBoundAnalysis}. For any dictionary~$\Phi$, the success probability $\epsilon(x,\Phi)$ can be expressed in terms of a union of events
\begin{equation}\label{UnionProbability}
	\prob{\epsilon(x,\Phi) \leq \varepsilon } = \prob{\bigcup\limits_{i=1}^{N} \norm{\proj{x}{\psi_i}}^2 \geq 1-\varepsilon^2},
\end{equation}
where \( \{\psi_i\}_{i=1}^N \) are all the $ N = \binom{n}{k} $ subspaces that are spanned by some choice of \(k\) atoms from \(\Phi\). Applying the union bound on \eqref{UnionProbability}, we get
\begin{equation}\label{UnionBoundProb}
	\prob{ \bigcup\limits_{i=1}^{N} \norm{ \proj{x}{\psi_i} }^2 \geq 1-\varepsilon^2 } \leq \sum\limits_{i=1}^{N} \prob{ \norm{ \proj{x}{\psi_i} }^2 \geq 1-\varepsilon^2 }.
\end{equation}
From Lemma~\ref{RandomSubspaceProjectionTheorem}, we know that \(\norm{\proj{x}{\psi_i}}^2 \thicksim \myBetai{\frac{k}{2}}{\frac{d-k}{2}} \) for all \(i\). Thus, using the inequality from \eqref{BetaUpperBound},\eqref{BetaUpperBound2} we have that
\begin{equation}\label{eq:UnionBoundRandom1}
	\prob{\bigcup\limits_{i=1}^{N} \norm{\proj{x}{\psi_i}}^2 \geq 1-\varepsilon^2} \leq \binom{n}{k} \exp\left\{-\frac{d}{2} \relent{1-\sparsity}{\varepsilon^2}\right\}.
\end{equation}
To bound the binomial coefficient, we use the next lemma from \cite{ash1965information}.
\newtheorem{BinomialCoefficientBounds}[Lemmas]{Lemma}
\begin{BinomialCoefficientBounds}[\cite{ash1965information}] \label{BinomialCoefficientBoundsTheorem}
	Let \( n,k \in \N \) such that \(  0< \frac{k}{n} <1 \) then
	\begin{equation}
		\frac{\exp\{n  \entropy{\frac{k}{n}} \} }{ \sqrt{ 8 k (1 - \frac{k}{n} ) } } \leq \binom{n}{k} \leq \frac{\exp\{ n \entropy{\frac{k}{n}} \}}{ \sqrt{ 2 \pi k (1 - \frac{k}{n} ) } },
	\end{equation}
	where \(\entropyi{\alpha}\) is the entropy of the \(\text{Bernoulli}(\alpha)\) distribution, defined as
	\begin{equation}\label{Entropy}
		\entropy{\alpha} = -\alpha\log(\alpha) - (1-\alpha)\log(1-\alpha).
	\end{equation}
\end{BinomialCoefficientBounds}
This bound on the binomial coefficient gives a slightly better result than the bound \eqref{BinomialUpperBound}. Using this lemma in \eqref{eq:UnionBoundRandom1}, we obtain
\begin{equation}
	\prob{\epsilon(x,\Phi) \leq \varepsilon }  \leq \frac{	
	\exp\left\{ - \frac{d}{2} \relent{1-s}{\varepsilon^2} + n \entropy{ \frac{k}{n} } \right\} }
	{ \sqrt{2\pi k(1 - \frac{k}{n})}}.
\end{equation}
The right hand side of this inequality is independent of the dictionary \( \Phi \). Therefore, by taking the maximum over all dictionaries \( \Phi \in \R ^{ d \times n } \) we get
\begin{equation}\label{AverageCaseLowerBoundResult}
	\optSuccProbWC  \leq \frac{	
	\exp\{ - d(\frac{1}{2} \relent{1-s}{\varepsilon^2} - \overcomp \,\entropy{ \frac{\sparsity}{\overcomp} }) \} }
	{ \sqrt{2\pi  d \sparsity (1 - \frac{\sparsity}{\overcomp})}},
\end{equation}
where we used~\eqref{CardinalParametersDef}. This proves the second statement of Theorem~\ref{AverageCaseLowerBoundTheorem}.

To prove the first part, let us derive a sufficient condition for the argument of the exponent in~\eqref{AverageCaseLowerBoundResult} to be negative. It is easy to show that \( -(1-\alpha)\log(1-\alpha) \leq \alpha  \) for all \( \alpha \in [0,1] \). Hence \( \overcomp \entropyi{\frac{\sparsity}{\overcomp}} \leq -\sparsity \log(\frac{\sparsity}{\overcomp}) +\sparsity \), which implies that
\begin{equation}
	\frac{1}{2} \relent{1-s}{\varepsilon^2} - \overcomp \,\entropy{ \frac{\sparsity}{\overcomp} } \geq \frac{1}{2} \relent{1-s}{\varepsilon^2} + \sparsity \log\left(\frac{\sparsity}{\overcomp}\right) -\sparsity .
\end{equation}
Therefore, to ensure that the left hand side is positive, we will require that \(  \frac{1}{2} \relent{1-s}{\varepsilon^2} + \sparsity \log(\frac{\sparsity}{\overcomp}) -\sparsity > 0 \). Isolating~$\overcomp$ leads to the condition
\begin{equation}\label{eq:UnionBoundRandom2}
	\overcomp \leq  e^{-1} \sparsity \exp\left\{\frac{1}{2\sparsity} \relent{1-\sparsity}{\varepsilon^2} \right\}.
\end{equation}
We thus conclude that when this condition holds,
\begin{equation}
	\lim\limits_{d \rightarrow \infty } \optSuccProbWC = 0.
\end{equation}
Substituting $\relent{1-\sparsity}{\varepsilon^2}$, the right-hand side of \eqref{eq:UnionBoundRandom2} coincides with the right-hand side of \eqref{AverageCaseLowerBoundEq}, thus completing the proof of Theorem~\ref{AverageCaseLowerBoundTheorem}.

\subsection{Average-case upper bound}\label{AverageCaseUpperBoundAnalysis}
We next prove Theorem \ref{AverageCaseUpperBoundTheorem}, which provides an average-case upper bound on the required overcompleteness. Recall from Lemma \ref{WorstCaseDistributionLemma} that the worst case distribution is isotropic. Therefore, as in section \ref{AverageCaseLowerBoundAnalysis}, we will analyze the case where \( x \thicksim N(0,I_{d \times d}) \). Since the signal \( x \) is stochastic, it will be more convenient to use a stochastic dictionary as well. The next lemma shows that this does not change the probability of success.
\newtheorem{StochasticDictionary}[Lemmas]{Lemma}
\begin{StochasticDictionary}[Stochastic dictionary] \label{StochasticDictionaryLemma}
	Let \( x \thicksim \Omega\) be a $d$-dimensional random vector and denote by \( \mathcal{D} (\R ^{ d \times n }) \) the collection of all distributions over \(\R ^{ d \times n }\). Then
	\begin{equation}
	\max_{\Phi \in \R ^{ d \times n } } \prob{ \epsilon(x,\Phi) \leq \varepsilon} = \max_{ \Theta \in \mathcal{D} (\R ^{ d \times n }) } \prob{ \epsilon(x,\Psi) \leq \varepsilon },
	\end{equation}
where \( \Psi\) in the right-hand side is a random dictionary with distribution $\Theta \in \mathcal{D} (\R ^{ d \times n })$, independent of \(x\).
\end{StochasticDictionary}
Here, the probability in the left-hand side is over the randomness of $x$ alone while the probability in right-hand side is over the randomness of both $x$ and $\Psi$.

Similarly to Sec.~\ref{WorstCaseUpperBoundAnalysis}, to obtain an upper bound on the required overcompleteness, we introduce two sub-optimal restrictions. First, instead of searching for the distribution of the dictionary that maximizes the success probability, we choose one particular distribution. Specifically, we focus on a block-diagonal dictionary of the form~\eqref{MyDictionary} with i.i.d.~entries distributed as $N(0,1)$. This choice is made mainly for convenience, and is worse than \eg a full Gaussian dictionary (whose atoms are spread isotropically). Second, we require the sparse representation to contain exactly one atom from each of the $k$ sub-dictionaries. Clearly, this latter limitation can only increase the representation error as it reduces the number of subspaces from \( \binom{n}{k} \) to \( (\frac{n}{k})^k \leq \binom{n}{k} \). The advantage of this limitation is that, as noted in Sec.~\ref{WorstCaseUpperBoundAnalysis}, it makes the squared representation error~\eqref{RepresentationErrorEq} separable (see~\eqref{ModificationInequality}). This allows us to bound the normalized $k$-sparse representation error of any signal $x$ over our block-diagonal $\Phi$ by
\begin{equation}\label{ModificationInequality2}
	\epsilon^2(x,\Phi) \leq \frac{1}{\norm{x}^2 } \sum_{i=1}^{k} \min_{1 \leq j \leq m } \norm{x_i - \proj{x_i}{\varphi^{(i)}_j}}^2,
\end{equation}
where we used the notation \eqref{MySignal} for the partitioning of the signal into $k$ parts.

\begin{figure}[!t]
	\centering
	\includegraphics[width=3.5in]{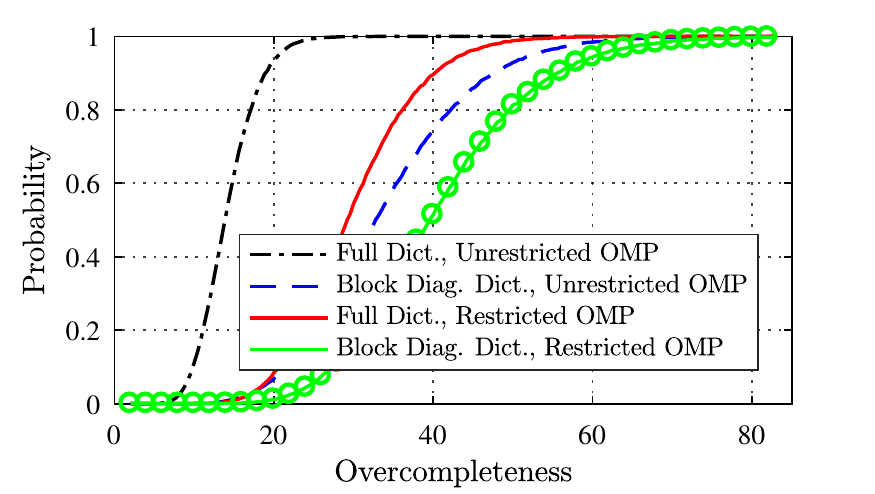}
	\caption{Probability of succeeding to sparsely represent a white Gaussian signal using the OMP algorithm, as a function of the overcompleteness of the dictionary. Here, the dimension is $d = 100$, the sparsity factor is $\sparsity = 0.2$, and the allowed reconstruction error corresponds to $\SNR = 10 [\dB] $. The black dotted curve shows the performance of standard OMP over a full Gaussian dictionary. The red solid curve is with a full Gaussian dictionary but with choice limitation. The blue dashed line is with a block diagonal Gaussian dictionary and without choice limitation. The green circled curve is with a block diagonal Gaussian dictionary and with choice limitation. We focus on lower bounding the green curve, thus also bounding the best achievable performance.}
	\label{fig:Modifications}
\end{figure}

Figure~\ref{fig:Modifications} illustrates the effect of these restrictions on the probability of success, when using the OMP algorithm to obtain the sparse representation. As can be seen, compared to using a full Gaussian dictionary with no restriction on the atom selection (black curve), each of these modifications introduces only a moderate increase in the required overcompleteness (red and blue curves). The combination of the two restrictions together, introduces an additional minor increase in the overcompleteness (green curve). Our approach is to lower bound the green curve.

Since $\normi{x_i- \proj{x_i}{\varphi^{(i)}_j}}^2 = \normi{x_i}^2- \normi{\proj{x_i}{\varphi^{(i)}_j}}^2$ and $\sum_{i=1}^k \normi{x_i}^2=\normi{x}^2$, the inequality in \eqref{ModificationInequality2} can be written as
\begin{equation}\label{UpdatedRepresentationError}
	\epsilon^2(x,\Phi) \leq 1-\frac{1}{\norm{x}^2} \sum_{i=1}^{k} \max_{1 \leq j \leq m } \norm{\proj{x_i}{\varphi^{(i)}_j}}^2.
\end{equation}
For simplicity, let us denote
\begin{equation}\label{RandomVariableDefinition}
	\gamma_i = \frac{\norm{x_i}^2}{\norm{x}^2} , \quad y^{(i)}_j = \frac{\norm{ \proj{x_i}{\varphi^{(i)}_j}}^2} {\norm{x_i}^2}, \quad Z_i = \max_{1 \leq j \leq m } y^{(i)}_j.
\end{equation}
Then we can rewrite \eqref{UpdatedRepresentationError} as
\begin{equation}\label{eq:gammai_Zi}
\epsilon^2(x,\Phi) \leq 1-\sum_{i=1}^{k} \gamma_i Z_i.
\end{equation}
Thus, we have the following lower bound on the optimal success probability
\begin{equation}\label{ProbabilityInequality}
	\optSuccProbWC \geq  \prob{ \epsilon(x,\Phi) \leq \varepsilon } \geq \prob{ \sum_{i=1}^{k} \gamma_i Z_i  \geq 1- \varepsilon^2 },
\end{equation}
where the first inequality is because our $\Phi$ is block diagonal and the second inequality follows from  \eqref{eq:gammai_Zi}, which is due to the atom choice selection. Our goal is to show that the series $\sum_{i=1}^{k} \gamma_i Z_i$ converges to its mean as $k\to\infty$, so that if its mean is greater than $1- \varepsilon^2$, then $\optSuccProbWC\to 1$. We will show this by explicitly calculating the mean and variance of the series.

From Lemma~\ref{RandomSubspaceProjectionTheorem}, we know that $  y^{(i)}_j \thicksim \myBetai{\frac{1}{2}}{\frac{1-\sparsity}{2\sparsity}}$ for all $i,j$, and \( \gamma_i \thicksim \myBetai{\frac{d}{2k}}{\frac{d(k-1)}{2k}} \) for all $i$. Using the properties of the Beta distribution, this implies that
\begin{equation}\label{MeanAndVarianceGamma}
	\Ei{\gamma_i} = \frac{1}{k}, \qquad \Var(\gamma_i) = \frac{2}{k^2} \frac{k-1}{d+2}.
\end{equation}
Since the random variables $\{ y^{(i)}_j : i \leq k , j \leq m \}$ are identically distributed, so are \( \{Z_i\}\). Furthermore, it can be shown that \( \{ y^{(i)}_j : i \leq k , j \leq m \} \) are mutually independent and are also independent of \( \{\gamma_i\}\) (see proof in Appendix \ref{DistributionPropertiesProof}). This implies that the random variables \( \{Z_i\}\) are also mutually independent and are independent of \( \{\gamma_i\}\).

Let us denote
\begin{equation}\label{MeanAndVarianceZ}
	\E{Z_i} = \mu, \qquad \Var(Z_i) = \sigma ^2.
\end{equation}
Then
\begin{equation}\label{SeriesExpectation}
	\E{\sum_{i=1}^{k} \gamma_i Z_i} = \sum_{i=1}^{k}  \E{\gamma_i} \E{Z_i} = \sum_{i=1}^{k}  \frac{1}{k} \mu = \mu.
\end{equation}
Furthermore,
\begin{align}
	\Var\left(\sum_{i=1}^{k} \gamma_i Z_i\right) &= \E {\left( \sum_{i=1}^{k} \gamma_i Z_i-\mu \right)^2} \nonumber\\
    &= \E{\left( \sum_{i=1}^{k} \gamma_i (Z_i-\mu) \right)^2 } \nonumber\\
    &= \E{\sum_{i=1}^{k} \gamma_i^2 (Z_i-\mu)^2} \nonumber\\
    &= \sum_{i=1}^{k} \E{\gamma_i^2} \E{(Z_i-\mu)^2},
\end{align}
where the second equality follows from the fact that \(\sum_{i=1}^{k} \gamma_i = 1\) and the third and fourth equalities follow from the independence of $\{Z_i\}$ and $\{\gamma_i\}$, which implies that \( \E{\gamma_i \gamma_j (Z_i-\mu)(Z_j-\mu)} = \E{\gamma_i \gamma_j} \E{Z_i-\mu}\E{Z_j-\mu}\) for all \( i \neq j \) (and obviously \( \E{Z_i-\mu}=0 \) for all $i$). Using the second moment formula \(\E{\gamma_i^2} = \Var(\gamma_i) + (\E{\gamma_i})^2\), with the variance and expectation given in \eqref{MeanAndVarianceGamma}, and writing \(\E{(Z_i-\mu)^2} = \sigma^2\), we can further simplify this expression as
\begin{align}\label{SeriesVariance}
	\Var\left(\sum_{i=1}^{k} \gamma_i Z_i\right)
    &= \sigma^2\sum_{i=1}^k \left(\Var(\gamma_i) + (\E{\gamma_i})^2\right) \nonumber\\
    &= k\sigma^2\left(\frac{1}{k^2}+\frac{2}{k^2} \frac{k-1}{d+2}\right) \nonumber\\
    &\leq \frac{(1+2\sparsity)\sigma^2}{k}.
\end{align}
We see that the expectation does not depend on $k$, whereas the variance decays as \( \frac{1}{k} \) (equivalently $\frac{1}{d}$). Therefore, from Chebyshev's inequality, this series converges in probability to its mean. Consequently, in order for the optimal success probability to converge to $1$ we must require that \( \mu \geq 1- \varepsilon^2 \).

Since $Z_i$ is the maximum over i.i.d variables, to bound its expectation $\mu$ we will use the next lemma \cite[Sec.~4.5]{david2004order}.
\newtheorem{MaxLowerBound}[Lemmas]{Lemma}
\begin{MaxLowerBound}[Lower bound on the order statistics of RVs] \label{MaxLowerBoundLemma}
	Let \( \{q_j\}_{j=1}^m \) be a set of i.i.d random variables defined on an interval \( \mathcal{I} \subseteq \R \), with marginal cumulative distribution function \( F_q(\alpha) \). Assume that \( F_q \) is differentiable, concave and strictly monotonically increasing on \( \mathcal{I} \). Let \( q_{[i]} \) denote the \(i\)th smallest value among \( \{q_j\}_{j=1}^m \), then
	\begin{equation}
		\Ei{q_{[i]}} \geq F_q^{-1} \left(\frac{i}{m+1}\right),
	\end{equation}
	where \( F_q^{-1}\) is the inverse function of \(F_q\) on the interval \( \mathcal{I}\) (satisfying \( F_q^{-1}(F_q(\alpha)) = \alpha \quad \forall \alpha \in \mathcal{I} \)).
\end{MaxLowerBound}
As mentioned above, the random variables $\{y_j^{(i)}\}$ are i.i.d and distributed as $\myBetai{\frac{1}{2}}{\frac{1-\sparsity}{2\sparsity}}$. Their cumulative distribution function \( F_y(\alpha) = \Ii{\alpha} {\frac{1}{2}} {\frac{1-\sparsity}{2\sparsity}} \) is differentiable, concave and strictly monotonically increasing. Hence, the conditions of Lemma~\ref{MaxLowerBoundLemma} are satisfied and we have that
\begin{equation}\label{ExpectationBound}
	\mu=\Ei{Z_i}= \E{\max_{1 \leq j \leq m } y^{(i)}_j} \geq I^{-1}_{\frac{m}{m+1}} \left( \frac{1}{2},\frac{1-\sparsity}{2\sparsity} \right) .
\end{equation}
For any given $\delta>0$, let us require that
\begin{align}\label{eq:expect_one_minus_delta}
I^{-1}_{\frac{m}{m+1}}\left(\frac{1}{2},\frac{1-\sparsity}{2\sparsity}\right)\geq 1-\delta^2,
 \end{align}
so that from \eqref{ExpectationBound}, we have that $\mu\geq1-\delta^2$. In particular, if we set $\delta=\varepsilon$, then we get $\mu\geq1-\varepsilon^2$, as desired. Due to the monotonicity of the function $I_{\frac{m}{m+1}}(\frac{1}{2},\frac{1-\sparsity}{2\sparsity})$ w.r.t.\@ $m$, the requirement \eqref{eq:expect_one_minus_delta} translates into the condition
\begin{equation} \label{SubDictionaryConnection}
	m \geq \frac{\I {1-\delta^2} {\frac{1}{2}} {\frac{1-\sparsity}{2\sparsity}}} { \I{\delta^2}{\frac{1-\sparsity}{2\sparsity}}{\frac{1}{2}}}
	= \frac{ \prob{ y_j^{(i)} \leq 1-\delta^2 } } {\prob{ y_j^{(i)} \geq 1-\delta^2 } }.
\end{equation}
Since $\overcomp=n/d=mk/d=m\sparsity$, we conclude that if the overcompleteness ratio satisfies
\begin{equation}\label{eq:overcomp_Idelta}
	\overcomp \geq \frac{ \sparsity } { \I{\delta^2}{\frac{1-\sparsity}{2\sparsity}}{\frac{1}{2}} },
\end{equation}
then we are guaranteed to have $ \mu \geq 1-\delta^2$. Using \( \delta =\varepsilon \) and lower-bounding the denominator using Lemma \ref{BetaLowerBoundLemma}, we get that if
\begin{equation}
	\overcomp \geq \sqrt{\frac{\pi}{2} (1-\varepsilon^2 \frac{1-\sparsity}{1+\sparsity} ) } \sparsity^{\frac{1}{2}}  \left(\frac{1}{\varepsilon}\right)^{ \sparsity^{-1} - 1 },
\end{equation}
then \(\lim\limits_{d \rightarrow \infty } \optSuccProbWC = 1\), which proves the first part of Theorem \ref{AverageCaseUpperBoundTheorem}.

Next, we prove the second part of the theorem, which requires bounding the probability \( \probi{ \sum_{i=1}^{k} \gamma_i Z_i  \geq 1- \varepsilon^2 } \) from below for any finite dimension \( d \). We will assume from this point on that \eqref{eq:overcomp_Idelta} holds with some \( \delta < \varepsilon \), which ensures that $\mu \geq 1-\delta^2 > 1-\varepsilon^2$. Note that \(  \{ \gamma_i \} \) are not independent, as \( \sum_{i=1}^{k} \gamma_i = 1\) w.p.~1. Therefore, we cannot use bounds such as Hoeffding's inequality. Instead, we will use Cantelli's inequality, which can be seen as a one-sided Chebyshev inequality. Cantelli's inequality states that for any random variable \(W\),
\begin{equation}\label{GeneralCantelliInequality}
	\prob{W-\E{W} \geq \lambda } ~
	\begin{cases}
	\leq \frac{\Var(W)}{\Var(W) + \lambda^2} ,& \lambda > 0,\\
	\geq 1-\frac{\Var(W)}{\Var(W) + \lambda^2} ,& \lambda < 0.
	\end{cases}
\end{equation}
In our case, this inequality gives
\begin{align}\label{CantelliInequalitySimple}
	&\prob{\sum_{i=1}^{k} \gamma_i Z_i \geq 1-\varepsilon^2 } \nonumber\\
	&\hspace{2.5cm}= \prob{\sum_{i=1}^{k} \gamma_i Z_i - \mu \geq -(\mu -1+\varepsilon^2) } \nonumber\\
	&\hspace{2.5cm}\geq  1-\frac{\Var(\sum_{i=1}^{k} \gamma_i Z_i)}{\Var(\sum_{i=1}^{k} \gamma_i Z_i) + (\mu -1+\varepsilon^2)^2} \nonumber\\
	&\hspace{2.5cm}= \left(1+\frac{\Var(\sum_{i=1}^{k} \gamma_i Z_i)}{(\mu -1+\varepsilon^2)^2}\right)^{-1}.
\end{align}
Thus, using \eqref{SeriesVariance} and writing $k=\sparsity d$, we get from \eqref{ProbabilityInequality} that
\begin{equation}\label{NumericStochasticBound}
	\optSuccProbWC \geq \left(1+\frac{(1+2\sparsity)\sigma^2}{(\mu -1+\varepsilon^2)^2 \sparsity} \frac{1}{d} \right)^{-1}.
\end{equation}
We note that this bound can be computed by evaluating the terms \(\mu\) and \(\sigma^2\) numerically, either by numerical integration or by using Monte Carlo simulations.

To obtain an expression that does not require numerical approximations, we can replace $\mu$ in \eqref{NumericStochasticBound} by its lower bound $1-\delta^2$ and also replace $\sigma$ by an upper bound as follows (see proof in Appendix \ref{VarianceUpperBoundProof}).
\newtheorem{VarianceBound}[Lemmas]{Lemma}
\begin{VarianceBound}[Upper bound on the variance] \label{VarianceUpperBoundLemma}
	Let \( W \) be a random variable defined on the interval \( [0,1] \) with cumulative distribution function \( F_W \). Then for every \( \rho \in [0,\frac{1}{2}] \),
	\begin{equation}
		\Var(W) \leq (1-2\rho)F_W(1-2\rho)+\rho^2.
	\end{equation}
\end{VarianceBound}
In our case, \( F_{Z_i}(\alpha) = ( \I{\alpha}{\frac{1}{2}}{\frac{1-\sparsity} {2\sparsity}})^m \). Thus, by setting \( \rho = \varepsilon^2 \leq \frac{1}{2} \) we have
\begin{equation} \label{VarBound}
	\sigma^2 = \Var(Z_i) \leq (1-2\varepsilon^2)( \I{1-2\varepsilon^2} {\tfrac{1}{2}}{\tfrac{1-\sparsity} {2\sparsity}})^m+\varepsilon^4.
\end{equation}
Remark: There exist several bounds on the variance of a bounded random variable, the most popular of which are Popoviciu's inequality and the Bhatia-Davis inequality. Popoviciu's inequality uses no knowledge on the probability distribution besides its support, and thus only gives \( \sigma^2\leq \frac{1}{4} \) in our case. The Bhatia-Davis inequality relies on the additional knowledge of the expectation and thus gives the slightly better result \(\sigma^2\leq(1-\mu )\mu \) in our case. However, considering that \( 1-\mu \simeq \varepsilon^2 \) this bound is only on the order of $\varepsilon^2$. In Lemma \ref{VarianceUpperBoundLemma} we also assume to have the cumulative distribution function of the random variable. This gives us an upper bound that scales like \( \varepsilon^4 \), as the left term in \eqref{VarBound} can become arbitrarily small when $m$ is large.

Using \eqref{VarBound} and $\mu>1-\delta^2$ in \eqref{NumericStochasticBound}, gives
\begin{equation}\label{FinalStochasticBound}
	\optSuccProbWC \geq \left( 1 + \frac{ (1-2\varepsilon^2) (\I{1-2\varepsilon^2} {\frac{1}{2}}{\frac{1-\sparsity}{2\sparsity}})^m+\varepsilon^4} {(\varepsilon^2-\delta^2)^2} \frac{(1+2\sparsity)}{\sparsity d}\right)^{-1},
\end{equation}
which proves the second statement of Theorem \ref{AverageCaseUpperBoundTheorem}.

\section{Numerical Simulations} \label{NumericalComp}

\begin{figure*}[!t]
	\centering
	\subfloat[Lower bound \eqref{WorstCaseLowerBoundEq},\eqref{AverageCaseLowerBoundEq}.]{
		\includegraphics[height=2in]{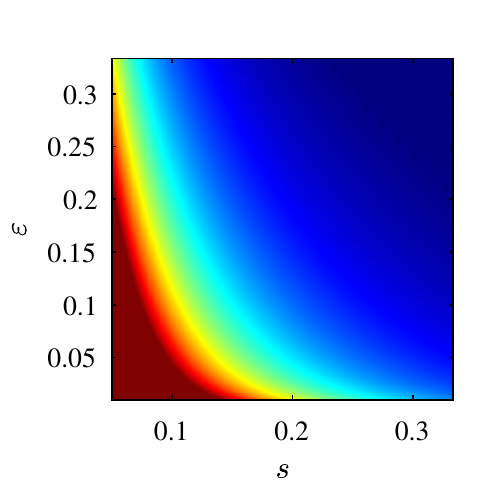}
		\label{Deterministic_Necessary_fig}
	}
	\subfloat[Average-case upper bound \eqref{AverageCaseUpperBoundEq}.]{
		\includegraphics[height=2in]{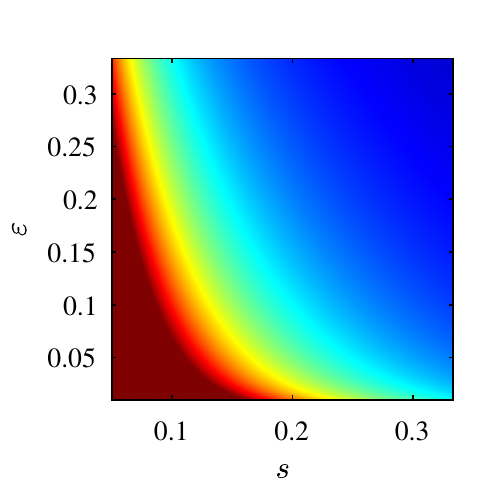}
		\label{Deterministic_Sufficient_fig}
	}
	\subfloat[Worst-case upper bound \eqref{WorstCaseUpperBoundEq}.]{
		\includegraphics[height=2in]{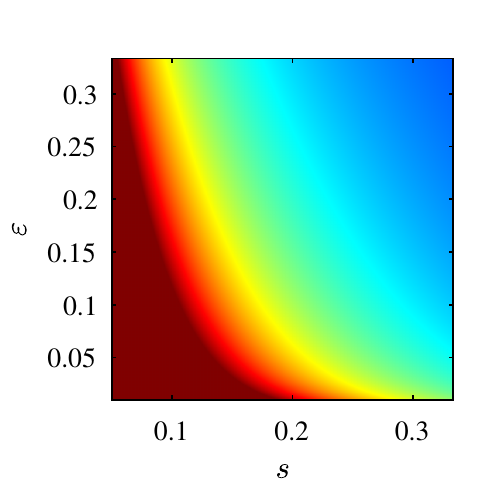}
		\label{Stochastic_Sufficient_fig}
	}
	\subfloat{
		\includegraphics[height=2in,trim={0.15in 0 0.3in 0},clip]{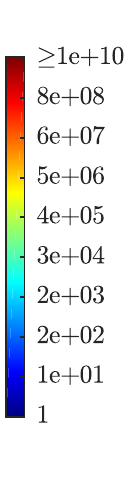}
		\label{colorbar}
	}
	\caption{Our bounds on the minimal overcompleteness that allows universal sparse representation, as functions of the sparsity factor $\sparsity$ and the permissible normalized error $\varepsilon$. Note that the color scale is logarithmic.}
	\label{Surfaces_fig}
\end{figure*}

In this section, we present simulations that demonstrate the bounds from Section \ref{MainResults}. Figure~\ref{Surfaces_fig} depicts the values of the worst-case and average-case lower-bounds \eqref{WorstCaseLowerBoundEq},\eqref{AverageCaseLowerBoundEq}, the worst-case upper bound \eqref{WorstCaseUpperBoundEq} and the average-case upper bound \eqref{AverageCaseUpperBoundEq}, as functions of the allowed error \( \varepsilon \) and the sparsity \( \sparsity \). Note that the color scale in all plots is logarithmic, thus highlighting the fact that the minimal required overcompleteness becomes extremely large for small values of \( \varepsilon \) and \( \sparsity \). We can see the asymmetrical dependency of the overcompleteness on \( \sparsity \) and \( \varepsilon \), which is exponential in \( \sparsity^{-1} \) and only polynomial in \( \varepsilon^{-1}  \). This illustrates that for small values of $\sparsity$, it is practically impossible to achieve universal sparse representation with any reasonable error $\varepsilon$ (the required overcompleteness is extremely large). However, for small values of $\varepsilon$, universal sparse representation may still be practical if the sparsity is not too small (\eg $\sparsity\approx 0.3$).

\begin{figure}[!t]
	\centering
	\includegraphics[width=3.5in]{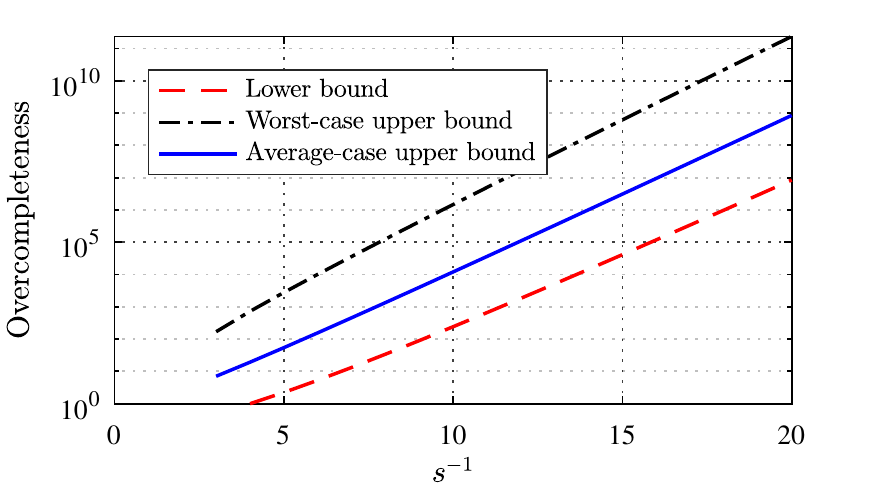}
	\caption{Overcompleteness as a function of \( \sparsity^{-1}\) at \(\SNR = 10\)dB.}
	\label{Overcomp_Spars_fig}
\end{figure}

\begin{figure}[!t]
	\centering
	\includegraphics[width=3.5in]{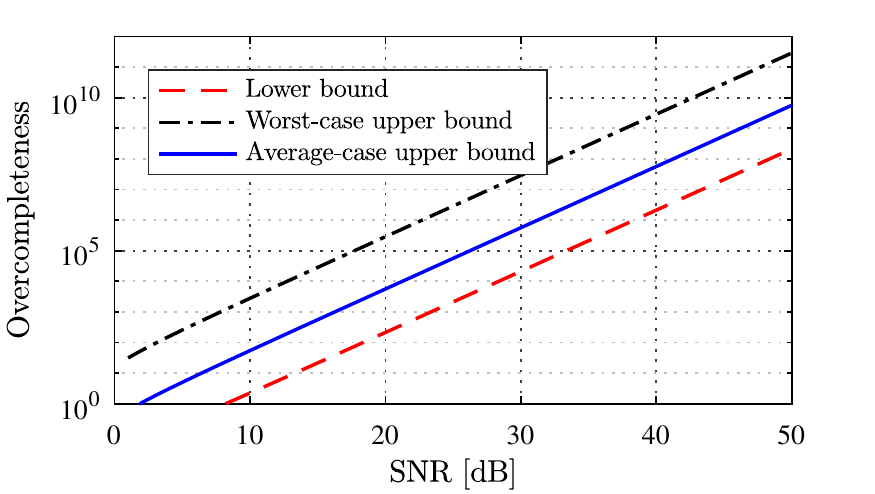}
	\caption{Overcompleteness as a function of \(\SNR\) at a sparsity of \( \sparsity=0.2 \).}
	\label{Overcomp_SNR_fig}
\end{figure}

To better visualize the differences between the bounds, Figs.~\ref{Overcomp_Spars_fig} and \ref{Overcomp_SNR_fig} show slices from the two-dimensional surfaces. Specifically, Fig.~\ref{Overcomp_Spars_fig} depicts the bounds as functions of \( \sparsity^{-1} \) at a constant representation error $\varepsilon$ corresponding to $\SNR_{\text{dB}} = 20\log_{10}(1/\varepsilon)=10$dB. Figure~\ref{Overcomp_SNR_fig} shows the bounds as functions of the SNR at a constant sparsity factor of $\sparsity=0.2$. Here we can see that the worst-case upper bound is quite pessimistic with respect to the average case one.

We next compare our bounds to the actual performance of a sparse coding algorithm. To the best of our knowledge, there exists no practical method for calculating the worst-case error \( \max_{x \in \R^d } \epsilon(x,\Phi) \) for a given dictionary $\Phi$. This means that we cannot verify whether a given $\Phi$ is a universal \(k\)-sparse representation dictionary. Consequently, we focus on examining only the average case scenario. In this setting, we take \(x \in \R^d \) to be a Gaussian vector with i.i.d coordinates, which, according to Lemma \ref{WorstCaseDistributionLemma}, is a worst case distribution. For any given $\Phi$, this allows us to easily approximate the probability of success $\prob{ \epsilon(x,\Phi) \leq \varepsilon }$, simply by applying the OMP algorithm on many draws of $x$ and counting the relative number of times the resulting sparse approximation satisfies our error constraint. This still leaves us with the problem of choosing the optimal $\Phi$. Since there is no closed form expression for the optimal $\Phi$, here we make a suboptimal choice, which is to take the dictionary to be a random matrix with Gaussian i.i.d entries. Recall that according to Lemma~\ref{StochasticDictionaryLemma}, the best possible probability of success is the same whether we restrict the search for deterministic dictionaries or also allow random dictionaries.

\begin{figure}[!t]
	\centering
	\includegraphics[width=3.5in]{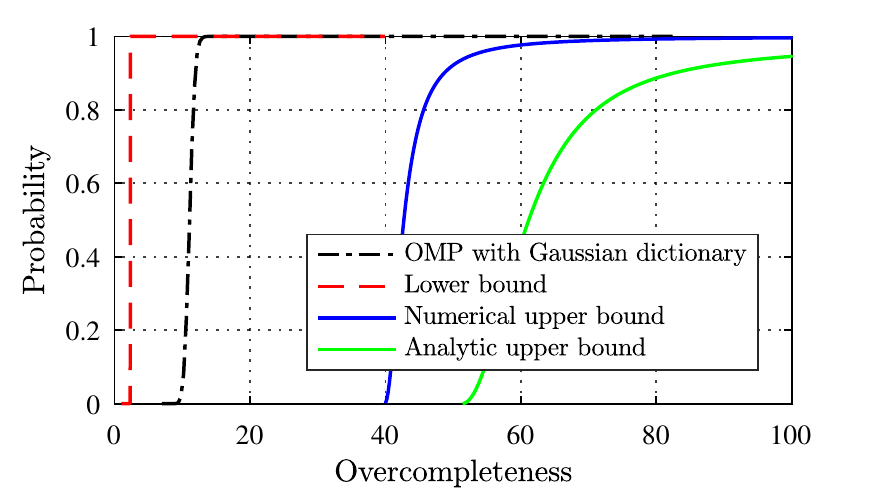}
	\caption{Probability of succeeding to sparsely represent a white Gaussian signal, as a function of the dictionary's overcompleteness. Here, the dimension is \( d = 1600\), the sparsity factor is $\sparsity = 0.2$, and the permissible error is  $\SNR = 10 {[\dB]}$.}
	\label{Prob_overcomp_fig}
\end{figure}

Figure~\ref{Prob_overcomp_fig} compares the probability of success of the OMP algorithm over a Gaussian dictionary to the lower and upper bounds on the probability of success \eqref{AverageCaseLowerBoundProb} and \eqref{AverageCaseUpperBoundProb2} as well as to the tighter bound \eqref{NumericStochasticBound}, which we calculated numerically. This simulation was carried out for a fixed dimension of $d=1600$. As can be seen, the success probability is indeed between the upper and lower bounds. Moreover, it exhibits a sharp phase-transition at some overcompleteness. Below this critical overcompleteness, the probability of success is nearly 0, while above the threshold it climbs very steeply towards 1. Obviously, both our choice of sparse-coding algorithm and our choice of dictionary are suboptimal. Therefore, we must keep in mind that the simulation gives us an underestimation of the optimal performance (the true best achievable probability of success is actually higher than the black dash-dotted curve).

\begin{figure}[!t]
	\centering
	\includegraphics[width=3.5in]{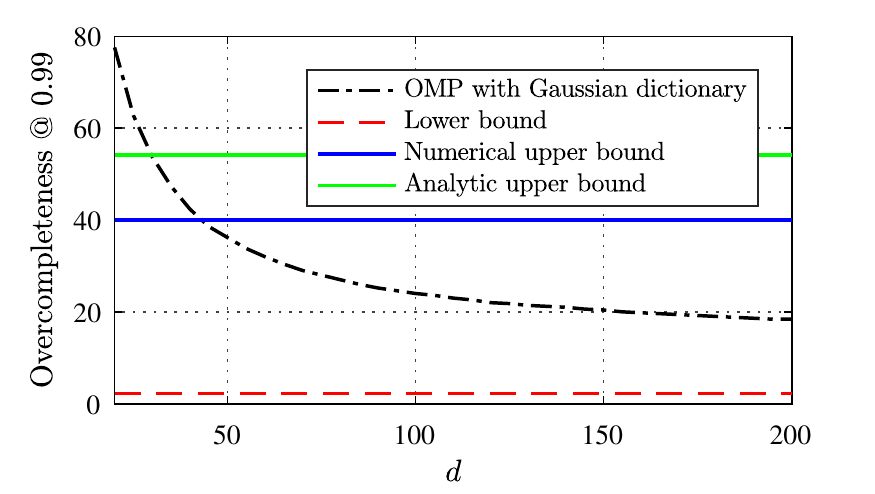}
	\caption{Minimal overcompleteness that allows to find a sparse representation for a white Gaussian signal with probability 0.99, as a function of the dimension $d$. Here, the sparsity factor is \( \sparsity = 0.2\) and the permissible normalized error is $\SNR = 10 {[\dB]} $.}
	\label{Overcomp_dimension_fig}
\end{figure}

Figure~\ref{Overcomp_dimension_fig} demonstrates the asymptotic behavior in Theorems~\ref{AverageCaseLowerBoundTheorem} and~\ref{AverageCaseUpperBoundTheorem} (\ie the bounds in \eqref{AverageCaseLowerBoundEq} and \eqref{AverageCaseUpperBoundEq}). Here, we compare the bounds to the minimal overcompleteness that allows to obtain a sparse representation with overwhelming probability. Specifically, we set a threshold of 0.99 on the success probability, and numerically found the minimal overcompleteness that allowed to surpass this success rate. This was done by gradually increasing the overcompleteness ratio, for each dimension $d$, until we hit the 0.99 success probability threshold for the first time. As can be seen, at high dimensions~$d$, the overcompleteness required for overwhelming success probability is indeed between the two bounds, as Theorems \ref{AverageCaseLowerBoundTheorem} and \ref{AverageCaseUpperBoundTheorem} predict.

\section{Conclusion}
In this paper, we presented and studied the \emph{universal sparse representation} problem, which relates to the ability of constructing sparse approximations to all signals in the space, up to a predefined error. We analyzed the problem in a deterministic setting as well as in a stochastic one. In both cases, we derived necessary and sufficient conditions on the minimal required overcompleteness. Our conditions have simple explicit forms, and, as we illustrated through simulations, accurately capture the behavior of sparse coding algorithms in practice.

\appendices

\section{Proof of Lemma \ref{BetaLowerBoundLemma}} \label{BetaBoundProof}
By the definition of the Beta distribution, we know that
\begin{equation}
	\I{x}{a}{b} = \frac{1}{B(a ,b)} \int_{0}^{x} t^{a-1}(1-t)^{b-1}dt,
\end{equation}
where \( B(a ,b) \) is the Beta function with parameters \(a\) and \(b\). We note that for \(  0<b \leq 1 \) the function \( w(t) = (1-t)^{b-1} \) is convex in \(t\). Moreover we have that
\begin{equation}
	 \int_{0}^{x} \frac{a}{x^a} t^{a-1}dt = 1.
\end{equation}
Let us represent the incomplete beta function as follows
\begin{equation}
	\I{x}{a}{b} =  \frac{x^a}{aB(a,b)} \int_{0}^{x} \frac{a}{x^a} t^{a-1}w(t)dt.
\end{equation}
Then by Jensen's inequality we have that
\begin{align}
	\int_{0}^{x} \frac{a}{x^a} t^{a-1}w(t)dt &\geq w\left( \int_{0}^{x} \frac{a}{x^a} t^{a-1}tdt  \right) \nonumber\\
	&= w\left(x \frac{a}{a+1}\right) \nonumber\\
	&= \left( 1-x \frac{a}{a+1}\right)^{b-1}.
\end{align}
Therefore
\begin{equation}
	\I{x}{a}{b} \geq \frac{x^a}{aB(a,b)(1-x\tfrac{a}{a+1})^{1-b}}.
\end{equation}
The Beta function can be expressed in terms of Gamma functions as
\begin{equation}
	B(a,b) = \frac{\Gamma(a)\Gamma(b)}{\Gamma(a+b)}.
\end{equation}
From the properties of the Gamma function, we know that \( a\Gamma(a) = \Gamma(a+1) \). Therefore,
\begin{equation}
	aB( a , b ) = \Gamma(b) \frac{ \Gamma(a+1)} {\Gamma(a+b)}.
\end{equation}
The ratio of the gamma functions is bounded by \cite[(3.2)]{mukhopadhyay2016stirling}
\begin{equation}
	\frac{ \Gamma(a+1)} {\Gamma(a+b)} < (a+b)^{1-b},
\end{equation}
which concludes the proof.

\section{Proof of Lemma \ref{WorstCaseDistributionLemma}} \label{WorstCaseDistributionProof}

To prove Lemma \ref{WorstCaseDistributionLemma}, we first prove the following useful property
\newtheorem{IsotropicDistributionLemma}[Lemmas]{Lemma}
\begin{IsotropicDistributionLemma}[Isotropic distribution]
	\label{IsotropicDistribution}
	Let \( \mathcal{D} (\R^{d})  \) denote the collection of all probability distributions defined on \(\R^{d}\). Then, for every distribution \(\Omega \in  \mathcal{D}(\R^{d})\) there exists an isotropic distribution \(\Theta \in  \mathcal{D}(\R^{d})\) for which the optimal success probability is not greater than the optimal success probability for \(\Omega \). That is
	\begin{equation}
	\optSuccProb{\Theta} \leq \optSuccProb{\Omega}.
	\end{equation}
\end{IsotropicDistributionLemma}
\begin{IEEEproof}[Proof of Lemma \ref{IsotropicDistribution}]
Let \(x \thicksim \Omega\) be some \(d\)-dimensional random vector. Denote by \(SO(d)\) the \emph{Special Orthogonal} group in \(\R^d\), which corresponds to all \(d\times d\) rotation matrices (satisfying \(R^TR=RR^T=I\) and \(\text{det}(R)=1\)). It is well known that there exists an invariant measure in \(SO(d)\) \cite[Sec.2]{leon2006statistical}. Therefore, let \( M \) be a matrix chosen uniformly at random from \(SO(d)\), and independent of \(x\). Then define
\begin{equation}
	y = Mx.
\end{equation}
It is easy to see that the distribution of \(y\), which we denote by \( \Theta \), is isotropic. We will show that the optimal success probability for the distribution \( \Theta \) is no larger than that of \( \Omega \). Using the law of total probability we have
\begin{equation} \label{TotalProb}
	\max_{\Phi \in \R^{d \times n}} \prob{\epsilon(y,\Phi) \leq \varepsilon} = \max_{\Phi \in \R^{d \times n}} \E{\condProb{\epsilon(Mx,\Phi) \leq \varepsilon}{M}},
\end{equation}
where the left hand side is the definition of the optimal success probability \( \optSuccProb{\Theta} \). Let us recall the definition of the representation error
\begin{equation} \label{ErrorDef}
	\epsilon(Mx,\Phi) = \min_{\alpha \in \R^n} \frac{\norm{Mx-\Phi \alpha} }{\norm{Mx}} \quad \text{s.t.} \quad \norm{\alpha}_0 \leq k.
\end{equation}
By the properties of orthogonal matrices, we know that \( \norm{Mx-\Phi \alpha} = \norm{x- M^T \Phi \alpha} \) and \( \norm{Mx} = \norm{x} \). Therefore \(\epsilon(Mx,\Phi) = \epsilon(x, M^T \Phi)\), so that
\begin{equation} \label{Orthogonal}
	\condProb{\epsilon(Mx,\Phi) \leq \varepsilon}{M} = \condProb{\epsilon(x, M^T \Phi) \leq \varepsilon}{M}.
\end{equation}
The expectation of a pointwise maximum is greater or equal to the maximum of the expectation, \textit{i.e.}
\begin{align} \label{MaxInequality}
	&\max_{\Phi \in \R^{d \times n}}  \E{\condProb{\epsilon(x, M^T \Phi) \leq \varepsilon}{M}} \leq \nonumber\\
	&\hspace{2cm}\E{\max_{\Phi \in \R^{d \times n}} \condProb{\epsilon(x, M^T \Phi) \leq \varepsilon }{M}}.
\end{align}
Recall that \( x \) and \( M \) are statistically independent. Therefore,
\begin{equation}
	\max_{\Phi \in \R^{d \times n}} \condProb{\epsilon(x, M^T \Phi) \leq \varepsilon}{M} = \max_{\Phi \in \R^{d \times n}} \prob{\epsilon(x,\Phi) \leq \varepsilon}.
\end{equation}
We notice that the right hand side is independent of \(M\). Thus, putting this result in \eqref{MaxInequality} we get
\begin{equation}
	\optSuccProb{\Theta} \leq  \max_{\Phi \in \R^{d \times n}} \prob{\epsilon(x,\Phi) \leq \varepsilon} = \optSuccProb{\Omega},
\end{equation}
which completes the proof.
\end{IEEEproof}

We will now show that the minimum of the optimal success probability over the set \( \mathcal{D}(\R^{d}) \) is attained by an isotropic distribution. It is always true that there exists a sequence of distributions \( \{ \Omega_n \}_{n=1}^\infty \subset  \mathcal{D} (\R^{d}) \) on which the optimal success probability converges to the minimal value. Mathematically,
\begin{equation}
	\lim\limits_{n \rightarrow \infty} \optSuccProb{\Omega_n} = \min_{\Omega \in  \mathcal{D} (\R^{d})} \optSuccProb{\Omega}.
\end{equation}
By Lemma \ref{IsotropicDistribution} we know that for each  \( n \in \mathbb{N} \) there exist an isotropic distribution \( \Theta_n \) such that \( \optSuccProb{\Theta_n} \leq \optSuccProb{\Omega_n} \). It is easy to see from the definition that all isotropic distributions have the same optimal success probability. Let \( \Theta \) be a given isotropic distribution, then we have that
\begin{equation}
	\forall n \in \mathbb{N} \qquad \optSuccProb{\Theta} = \optSuccProb{\Theta_n} \leq \optSuccProb{\Omega_n}.
\end{equation}
The left hand side is independent of \(n\), then by taking the limit we get that
\begin{equation}
	\optSuccProb{\Theta} \leq \lim\limits_{n \rightarrow \infty}  \optSuccProb{\Omega_n} = \min_{\Omega \in  \mathcal{D} (\R^{d})} \optSuccProb{\Omega}.
\end{equation}
We thus conclude that \( \Theta \) is a minimizer, which completes the proof.

\section{Proof of Lemma \ref{StochasticDictionaryLemma}} \label{StochasticDictionaryProof}

On the one hand, a deterministic dictionary is a special case of a stochastic dictionary. Hence, we have the inequality
\begin{equation}
	\max_{\Phi \in \R ^{ d \times n } } \prob{ \epsilon(x,\Phi) \leq \varepsilon} \leq \max_{ \Theta \in \mathcal{D} (\R ^{ d \times n }) } \prob{ \epsilon(x,\Psi) \leq \varepsilon }.
\end{equation}
On the other hand, by the law of total probability,
\begin{equation}
	\max_{ \Theta \in \mathcal{D} (\R ^{ d \times n }) } \prob{ \epsilon(x,\Psi) \leq \varepsilon } = \max_{ \Theta \in \mathcal{D} (\R ^{ d \times n }) } \E{\condProb{ \epsilon(x,\Psi) \leq \varepsilon}{\Psi }}.
\end{equation}
Obviously the maximum is greater or equal to the expectation, that is
\begin{equation}
	\E{\condProb{ \epsilon(x,\Psi) \leq \varepsilon}{\Psi}} \leq \max_{\Psi \in \R ^{ d \times n } } \prob{ \epsilon(x,\Psi) \leq \varepsilon}.
\end{equation}
Here, we used the fact that \( x \) and \( \Psi \) are statistically independent to remove the conditioning in the right-hand term. The above inequality is true for any distribution \( \Theta \in \mathcal{D} (\R ^{ d \times n }) \), and thus in particular it applies also to the maximum. Therefore, we obtain the opposite inequality
\begin{equation}
	 \max_{ \Theta \in \mathcal{D} (\R ^{ d \times n }) } \prob{ \epsilon(x,\Psi) \leq \varepsilon } \leq \max_{\Psi \in \R ^{ d \times n } } \prob{ \epsilon(x,\Psi) \leq \varepsilon},
\end{equation}
and the result follows.

\section{Proof of Lemma \ref{VarianceUpperBoundLemma} } \label{VarianceUpperBoundProof}

The variance of \(W\) is defined as
\begin{equation}
	\Var(W) = \E{(W-\E{W})^2}.
\end{equation}
Let us consider this formula as a mean square error (MSE) between \(W\) and its expectation. It is well known that the expectation achieves the minimum MSE over all constants. Thus, by replacing the expectation with \(1-\rho \) we can only increase the value of the MSE. Hence we have
\begin{equation}
	\E{(W-\E{W})^2} \leq \E{(W-1+\rho)^2}.
\end{equation}
Using the law of total expectation we get that for any \( \rho \in [0,\frac{1}{2}] \),
\begin{align}
	& \E{(W-1+\rho)^2} \nonumber\\
	& = \condE{ (W-1+\rho)^2 }{ W \leq 1-2\rho} \times \prob{W \leq 1-2\rho} \nonumber\\
	& + \condE{ (W-1+\rho)^2 }{ W > 1-2\rho}  \times \prob {W > 1-2\rho}.
\end{align}
It is easy to see that if \(W\in [0,1-2\rho] \) then
\begin{equation}
	(W-1+\rho)^2 \leq (1-\rho)^2
\end{equation}
and if \(W\in (1-2\rho,1] \) then
\begin{equation}
	(W-1+\rho)^2 \leq \rho^2.
\end{equation}
Therefore we have that
\begin{align}
	\E{(W-1+\rho)^2} \leq & (1-\rho)^2 \times \prob{W \leq 1-2\rho} \nonumber\\
	 &+  \rho^2 \times \prob {W > 1-2\rho}.
\end{align}
By definition, \( \prob{W \leq 1-2\rho} = F_W(1-2\rho) \) and \( \prob {W > 1-2\rho} = 1-F_W(1-2\rho)  \). Therefore, after simple algebra we get that
\begin{equation}
	\E{(W-1+\rho)^2} \leq (1-2\rho)F_W(1-2\rho)+\rho^2,
\end{equation}
which concludes the proof.

Remark: If we choose \( \rho = \frac{1}{2} \), then we get the value of Popoviciu's inequality, which in this case is equal to \( \frac{1}{4} \).

\section{Proof of independence property} \label{DistributionPropertiesProof}

To show that the set \( \{ y^{(i)}_l : i \leq k , l \leq m \} \) is mutually independent and is independent of the set \( \{\gamma_i\}_{i=1}^k  \), we will use the characteristic function. Recall the definitions of these sets,
\begin{equation}
	\gamma_i = \frac{\norm{x_i}^2}{\norm{x}^2}, \qquad y^{(i)}_j = \frac{\norm{ \proj{x_i}{\varphi^{(i)}_j}}^2} {\norm{x_i}^2}.
\end{equation}
For a general random vector \( W \in \R^n \), the characteristic function is defined by
\begin{equation}
	\bigchi_W(t) = \E{e^{jt^TW}} , \qquad t \in \R^n.
\end{equation}
where \( j = \sqrt{-1} \) is the unit imaginary number. Let \( Y \) and \(\Gamma \) be vector representations of the sets  \( \{ y^{(i)}_l : i \leq k , l \leq m \} \) and \( \{\gamma_i\}_{i=1}^k  \) respectively (concatenated in arbitrary order in each vector). Then,
\begin{equation}
	\bigchi_{(Y, \Gamma)}(t,s) = \E{e^{jt^TY+js^T\Gamma}}, \qquad t \in \R^n,\ s \in \R^k
\end{equation}
where \(n = km \) is the number of random variables in the vector \( Y \). Using the law of total expectation we have that
\begin{equation}
	\E{e^{jt^TY+js^T\Gamma}} = \E{\condE{e^{jt^TY+js^T\Gamma}}{x}}.
\end{equation}
Given \(x\), the variables \( \{\gamma_i\}_{i=1}^k  \) are determined deterministically. Additionally, since \( \{ \varphi^{(i)}_l : i \leq k , l \leq m \} \) are mutually independent, we have that \( \{ y^{(i)}_l : i \leq k , l \leq m \} \) are conditionally independent given \(x\). Therefore, we can write
\begin{equation}
	\E{\condE{e^{jt^TY+js^T\Gamma}}{x}} = \E{e^{js^T\Gamma} \prod_{i=1}^{N}\condE{e^{jt_iY_i}}{x}}.
\end{equation}
From Lemma~\ref{RandomSubspaceProjectionTheorem} we notice that the conditional distribution of \( Y_i \) given \( x \) is the same as the unconditioned distribution. Thus, we have
\begin{equation}
	\condE{e^{jt_iY_i}}{x} = \E{e^{jt_iY_i}} = \bigchi_{Y_i}(t_i).
\end{equation}
We point out that the right term is a deterministic function, therefore
\begin{equation}
	\bigchi_{(Y, \Gamma)}(t,s) = \E{e^{js^T\Gamma}} \prod_{i=1}^{N}\bigchi_{Y_i}(t_i) = \bigchi_{\Gamma}(s) \prod_{i=1}^{N}\bigchi_{Y_i}(t_i),
\end{equation}
which proves the desired independence.


\bibliographystyle{IEEEtran}
\bibliography{bibliography}


%\newpage

\begin{IEEEbiographynophoto}{Rotem~Mulayoff}
received the B.Sc. (Summa Cum Laude) degree in electrical engineering from the Technion-Israel Institute of Technology, Haifa, Israel, in 2016, where he is currently working toward the Ph.D. degree.

From 2013 to 2016, he worked in the field of signal processing and algorithms in RAFAEL Advanced Defense Systems LTD. Since 2016, he has been a Teaching Assistant with the Viterbi Faculty of Electrical Engineering, Technion.
His research interests include signal processing and machine learning.

Mr. Mulayoff is the recipient of the Freescale Prize for 2015, the Meyer Fellowship and the Cipers Award for 2016 and the Porat Award for 2018.

\end{IEEEbiographynophoto}

\begin{IEEEbiographynophoto}{Tomer~Michaeli}
received the B.Sc. degree (Summa Cum Laude) and the Ph.D. degree in electrical engineering in 2004 and 2012, respectively, both from the Technion--Israel Institute of Technology.
From 2000 to 2008, he was a Research Engineer at RAFAEL Research Laboratories, Israel Ministry of Defense, Haifa. From 2012 to 2015 he was a Postdoctoral Fellow at the Weizmann Institute of Science, Israel. Since 2015 he is an Assistant Professor at the Faculty of Electrical Engineering at the Technion. His research interests lie in the areas of Signal and Image Processing, Computer Vision, and Machine Learning.

Dr. Michaeli was awarded the Andrew and Erna Finci Viterbi Fellowship in 2008 and in 2012, the Jacobs-QUALCOMM Fellowship in 2010, the Sir Charles Clore Postdoctoral Fellowship in 2014, the Horev Fellowship for Outstanding Young Faculty at 2015, and the Alon Fellowship at 2017. He won the Jury Award in 2011, the Hershel Rich Technion Innovation award in 2012, the Vivian Konigsberg Award for Excellence in Teaching in 2011, and was a co-author of the paper that won the Best Student Paper Award at the IEEEI Conference 2012, Israel.
\end{IEEEbiographynophoto}


%\vfill


\end{document}